\documentclass[prl,twocolumn,superscriptaddress,footinbib]{revtex4}

\usepackage[dvipdfmx,hiresbb]{graphicx}
\usepackage{color}
\usepackage{enumerate}
\usepackage{epsfig,subfigure}
\usepackage{amsmath,amssymb,latexsym}
\usepackage{ascmac}
\usepackage{bm}
\usepackage{enumitem}
\usepackage{natbib}

\def\U#1{{\rm #1}} 

\newtheorem{theorem}{{\bf Theorem}}

\newtheorem{coro}{{\bf Corollary}}
\newcommand{\bra}[1]{\langle #1 |}
\newcommand{\ket}[1]{| #1 \rangle}

\newcommand{\expect}[1]{\left\langle #1 \right\rangle} 

\newcommand{\sq}{\qquad $\blacksquare$}

\newcommand{\no}{\notag}

\newcommand{\fin}{\U{fin}}
\newcommand{\leftbar}{\left|\left|}
\newcommand{\rightbar}{\right|\right|}
\newcommand{\ec}{\U{EC}}
\newcommand{\pa}{\U{PA}}
\newcommand{\emm}{\U{em}}
\newcommand{\code}{\U{code}}
\newcommand{\sample}{\U{samp}}

\newcommand{\rec}{\U{rec}}
\newcommand{\cnot}{\U{CNOT}}
\newcommand{\opt}{\U{opt}}

\def\Pr{\U{Pr}}
\def\tr{\U{tr}}

\def\vac{{\rm vac}}

\newcommand{\ph}{\U{ph}}
\newcommand{\bit}{\U{bit}}
\newcommand{\wt}{\U{wt}}

\begin{document}
\title
{Finite-key security analysis of differential-phase-shift quantum key distribution}
\author{Akihiro Mizutani}
\affiliation{Mitsubishi Electric Corporation, Information Technology R\&D Center, 5-1-1 Ofuna, Kamakura-shi, Kanagawa, 247-8501 Japan}
\affiliation{Faculty of Engineering, University of Toyama, Gofuku 3190, Toyama 930-8555, Japan}
\author{Yuki Takeuchi}
\affiliation{
NTT Communication Science Laboratories, NTT Corporation, 3-1 Morinosato Wakamiya, Atsugi, Kanagawa 243-0198, Japan
}
\author{Kiyoshi~Tamaki}
\affiliation{Faculty of Engineering, University of Toyama, Gofuku 3190, Toyama 930-8555, Japan}

\begin{abstract}
  {
Differential-phase-shift (DPS) quantum key distribution (QKD) is one of the major QKD protocols that can be implemented 
with a simple setup using a laser source and a passive detection unit. 
Recently, an information-theoretic security proof of this protocol has been established in [npj Quant. Inf. {\bf 5}, 87 (2019)] assuming 
the infinitely large number of emitted pulses. 
To implement the DPS protocol in a real-life world, it is indispensable to analyze the security with the finite number of emitted pulses. 
The extension of the security proof to the finite-size regime requires the accommodation of the statistical fluctuations to determine the amount of privacy amplification. In doing so, Azuma's inequality is often employed, but unfortunately we show that in the case of the DPS protocol, this results in a substantially low key rate. This low key rate is due to a loose estimation of the sum of probabilities regarding three-photon emission whose probability of occurrence is very small. The main contribution of our work is to show that this obstacle can be overcome by exploiting the recently found novel concentration inequality, Kato's inequality. As a result, the key rate of the DPS protocol is drastically improved. For instance, assuming typical experimental parameters, 
a 3 Mbit secret key can be generated over 77 km for 8.3 hours, which shows the feasibility of DPS QKD under a realistic setup.
} 
 \end{abstract}

\maketitle

\section{Introduction}
Quantum key distribution (QKD) realizes information-theoretically secure communication 
between two distant parties (Alice and Bob) against any eavesdropper (Eve). 
Since the first invention of the BB84 protocol~\cite{bb84}, various protocols have been proposed~\cite{e91,b92,six,sarg,cow,dpsPRL,inoyama2002,continuous}. 
Among them, the differential-phase-shift (DPS) protocol~\cite{inoyama2002} 
is considered to be one of the promising protocols for future QKD implementations. 
This is because the DPS protocol can be implemented with an experimentally simple setup using a laser source and a passive detection unit. 
The experimental demonstrations of this protocol have been conducted in~\cite{ex00,ex0,ex1} and also its field demonstration has been done in 
the Tokyo QKD network~\cite{ex2}. 
Also, security proofs of the DPS protocol have been intensively studied so far. 
In proving the security, the difficulty specific to this protocol is that one needs to deal with a very large Hilbert space since 
this protocol extracts sifted key information from the phase difference between adjacent pulses, and hence 
all the emitted pulses are continuously connected like a chain. 
To simplify the analysis, the previous security proofs have disentangled this chain by introducing a block. 
This block consists of some emitted pulses, and the protocol extracts only one sifted key bit from each block. 
For example, the first information-theoretic security proof~\cite{prl09} assumes that a single photon exists in each of the blocks. 
This impractical single-photon assumption has been mitigated to a block-wise phase-randomized coherent source in~\cite{tkk2012,mizu2017}. 
Applying a random phase shift to each block enables one to analyze the security for each photon number emission event separately. 
The recent work~\cite{npjmizu2018} has removed the need of the block-wise phase randomization and proven the security under 
simplified source assumptions including the case with two phase-modulated coherent states. 
Furthermore, the security proof in~\cite{mizutani_single} has extended the one in~\cite{npjmizu2018} 
to cover the case where the source emits any two identical and independent states. 
Importantly, these works~\cite{prl09,tkk2012,mizu2017,npjmizu2018,mizutani_single} guarantee the information-theoretic security of the DPS protocol, 
namely, these proofs are  valid under any of Eve's attack. 
Also, 
a recent work~\cite{eisei} has studied the performance of the DPS protocol by assuming a specific Eve's attack in the satellite environment.

The information-theoretic security proofs of the DPS protocol so far are only valid in the asymptotic regime, where the length of the sifted key is assumed to be infinite. 
Like other major QKD protocols~\cite{ncomm12,AL,m14,koashi19,npj21,azuma1,azuma2}, 
it is indispensable to reveal its key-generation efficiency with the finite-key length to implement the DPS protocol in real-life environments. 
In the finite-key analysis, the crucial is to evaluate statistical deviation terms of concentration inequalities 
in deriving an upper bound on the amount of privacy amplification. 
In so doing, it is important to employ an inequality that results in a small deviation with a smaller number of trials; 
otherwise the speed of convergence to the asymptotic key rate becomes slow, leading to a poor performance.

In this paper, we extend our previous information-theoretic security proof~\cite{npjmizu2018} of the DPS protocol to the finite-size one. 
As was implied in the previous work~\cite{npjmizu2018}, this extension 
can be achieved by using Azuma's inequality~\cite{Azuma} to deal with correlated random variables. 
This inequality is a well-known concentration inequality used in various security proofs~\cite{azuma0,azuma001,azuma01,azuma1,azuma12,azuma2}. 
Unfortunately, however, 
we reveal that the analysis with Azuma's inequality results in a substantially low key rate under a realistic experimental setup. 
To overcome this problem, we exploit Kato's inequality~\cite{kato}, which is the novel concentration inequality, 
and show that the key rate is drastically improved. 
More concretely, our numerical simulation shows that 
its achievable distance becomes more than three times longer than the one based on the analysis using Azuma's inequality 
(see Fig.~\ref{fig:comparison}). 
Note that using Kato's inequality instead of Azuma's one gives a significant improvement in the key rate only if the estimation of the leaked information 
involves events that occur with very small probability. 
This was pointed out in the recent finite-key analyses~\cite{npj21,sbull} of the twin-field QKD protocol. 
In our case, such a rare event is a detection event originating from emissions of three photons, and we show that its probability is small enough to enjoy the significant improvement with the use of Kato's inequality. 
We explain its details in Sec.~\ref{sec:upphase}.

The rest of the paper is structured as follows. 
Section~\ref{sec:ass} explains the assumptions we impose on the users' devices. 
We describe our DPS protocol in Sec.~\ref{pro} and prove its security in Sec.~\ref{sec:secu}. 
Section~\ref{sec:simu} presents the numerical simulation results of the key rate. 
In Sec.~\ref{sec:ad}, we compare the 
key rates obtainable using the analysis based on Kato's and Azuma's inequalities. 
Finally, Sec.~\ref{sec:conc} concludes our paper.

\section{Assumptions on devices}
\label{sec:ass}
Before describing the protocol, we summarize the assumptions we make on the source and measurement units. 
These are the same as those in our previous work~\cite{npjmizu2018}, but we describe them for the completeness of this paper. 
In this paper, we consider that Alice employs three pulses contained in a single block, and Alice and Bob try to extract a key bit from each 
block. 

\subsection{Assumptions on Alice's source unit}
\label{ass:A}
First, we list up the assumptions on Alice's source as follows. 

\begin{enumerate}[label=(A\arabic*)]
\item
\label{ASS1}
Alice randomly chooses a three-bit sequence $\bm{b}_A:=b^{(1)}_Ab^{(2)}_Ab^{(3)}_A\in\{0,1\}^3$, where 
bit $b^{(u)}_A$ is encoded only on the $u^{\U{th}}$ emitted pulse of system $S_u$. Depending on the chosen $\bm{b}_A$, 
Alice prepares the following three-pulse state of systems $\bm{S}:= S_1S_2S_3$:
\begin{align}
\hat{\rho}^{\bm{b}_A}_{\bm{S}}:=\bigotimes_{u=1}^3\hat{\rho}^{b^{(u)}_A}_{S_u}.
\end{align}
Here, $\hat{\rho}^{b^{(u)}_A}_{S_u}$ is a density operator of the $u^{\U{th}}$ pulse when $b^{(u)}_A$ is selected. 
We assume that the purified system $R_u$ of $\hat{\rho}^{b^{(u)}_A}_{S_u}$ is possessed by Alice, and Eve cannot access to system $R_u$. 
Note that state $\hat{\rho}^{b^{(u)}_A}_{S_u}$ is allowed to be different for each system $S_u$. 
\item
\label{assV}
The probability of the $u^{\U{th}}$ emitted pulse being the vacuum state is independent of bit $b^{(u)}_A$. 
That is, 
\begin{align}
\tr\left[\ket{\vac}\bra{\vac}\hat{\rho}^{0}_{S_u}\right]=\tr\left[\ket{\vac}\bra{\vac}\hat{\rho}^{1}_{S_u}\right]
\end{align}
holds for any $u$, where $\ket{\vac}$ denotes the vacuum state.
\item
\label{ASS3}
For any chosen bit sequence $\bm{b}_A$, 
the probability that any single block of pulses contains $n$ ($n\in\{1,2,3\}$) or more photons is upper-bounded by $q_n$: 
\begin{align}
\sum_{m\ge n}
\tr[\ket{m}\bra{m}\hat{\rho}^{\bm{b}_A}_{\bm{S}}]\le q_n,
\label{eq:qa}
\end{align}
where $\ket{m}$ denotes the photon-number state in all the optical modes. 
\end{enumerate}
Importantly, we do not assume block-wise phase randomization like in~\cite{tkk2012,mizu2017}. 
Note that such randomization enables us to regard the state of every single block as a classical mixture of the Fock states. 
However, our security proof holds without such an assumption and is valid even if there exists a phase coherence among the emitted blocks. 
This allows us to employ the source assumed in the original DPS protocol~\cite{dpsPRL}, 
which emits a pulse in a coherent state randomly chosen from $\{\ket{\alpha},\ket{-\alpha}\}$. 

We remark that the work~\cite{mizutani_single} has mitigated assumption~\ref{assV} to cover the case where Alice only knows 
the range of the probabilities of being the vacuum state. 
It could be possible to prove the security of the DPS protocol with this mitigated assumption in the finite-size regime, but for simplicity of discussion 
we adopt the above assumptions based on~\cite{npjmizu2018}.

\subsection{Assumptions on Bob's measurement unit}
Next, we explain the assumptions on Bob's measurement unit. 
\begin{enumerate}[label=(B\arabic*)]
\item
Bob measures incoming pulses using a one-bit delay Mach-Zehnder interferometer with 50:50 beam splitters (BSs). 
This delay is equal to the time interval of the neighboring emitted pulses. 
\item
The interfered pulses are detected by two photon-number-resolving (PNR) detectors, which discriminate the vacuum, 
a single photon, and two or more photons in a specific optical mode. 
We assume that the quantum efficiencies and dark countings are the same for both detectors. 
According to which PNR detector reports a click, Bob obtains a raw key bit $d\in\{0,1\}$. 
\end{enumerate}
For each incoming block, $j^{\U{th}}$ ($j\in\{1,2\}$) time slot is defined by the expected 
detection time where $j^{\U{th}}$ and $(j+1)^{\U{th}}$ incoming pulses interfere. 
Also, the $0^{\U{th}}$ and $3^{\U{rd}}$ time slots are defined by the expected detection time where the 
$1^{\U{st}} (3^{\U{rd}})$ incoming pulse and $3^{\U{rd}} (1^{\U{st}}$) one in the previous (next) block interfere.

\section{Actual protocol}
\label{pro}
\begin{figure}[t]
\includegraphics[width=8.5cm]{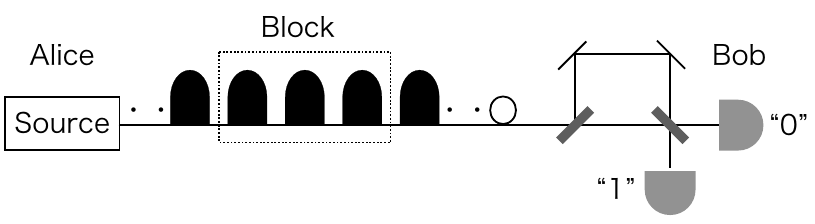}
\caption{
Schematics for our DPS protocol. Alice sends blocks of three pulses to Bob, and he 
receives them with the one-bit delay Mach-Zehnder interferometer and detectors. 
}
 \label{fig:setup}
\end{figure}
We describe our DPS protocol, which is the finite-size version of our previous protocol~\cite{npjmizu2018}. 
In its description, $\wt(\bm{b})$ denotes the number of 1's in a bit string $\bm{b}$. 
We depict a schematic diagram of our DPS protocol in Fig.~\ref{fig:setup}. 

\begin{enumerate}[label=(P\arabic*)]
\item
Alice and Bob respectively repeat the following procedures for $N_{\U{em}}$ rounds.
  \begin{enumerate}
\item
\label{pro:A}
Alice generates uniformly random three bits $\bm{b}_A\in\{0,1\}^3$ and sends three pulses in state $\hat{\rho}^{\bm{b}_A}_{\bm{S}}$ 
to Bob via a quantum channel. 
\item
Bob forwards the incoming three pulses into the Mach-Zehnder interferometer followed by photon detection by the PNR detectors.
We call the round {\it detected} if Bob detects exactly one photon in total among the 1$^{\U{st}}$ and 2$^{\U{nd}}$ time slots. 
The detection event at the $j^{\U{th}}$ ($j\in\{1,2\}$) time slot determines the raw key bit $d\in\{0,1\}$ depending on which of the two detectors clicks. 
\end{enumerate}
\item
\label{tt}
Bob takes note of a set of detected rounds $\mathcal{D}\subseteq\{1,...,N_{\emm}\}$ with length $N_{\det}:=|\mathcal{D}|$, 
a set of time slots $\bm{j}:=(j_i)_{i\in\mathcal{D}}$ and a raw key $\bm{d}:=(d_i)_{i\in\mathcal{D}}$. 
Here, $j_i$ and $d_i$ are $j$ and $d$ of the $i^{\U{th}}$ detected round, respectively. 
Bob associates each detected round with a {\it code} or {\it sample} round with probability $t$ or $1-t$, respectively with $0<t<1$. 
He defines the code set $\mathcal{D}_{\code}$ with length $N_{\code}:=|\mathcal{D}_{\code}|$, 
the sample one $\mathcal{D}_{\sample}:=\mathcal{D}\setminus\mathcal{D}_{\code}$ with length $N_{\sample}:=|\mathcal{D}_{\sample}|$, 
his sifted key $\kappa_B:=(d_i)_{i\in\mathcal{D}_{\code}}$ and the sample sequence $\kappa^{\sample}_B:=(d_i)_{i\in\mathcal{D}_{\sample}}$. 
\item
Bob announces $\mathcal{D}_{\code}$, $\mathcal{D}_{\sample}$, $\bm{j}$ and $\kappa^{\sample}_B$ to Alice 
through an authenticated public channel.
\item
\label{pr:bb}
Alice calculates her  sifted key $\kappa_A:=(b^{(j_i)}_A\oplus b^{(j_i+1)}_A)_{i\in\mathcal{D}_{\code}}$ and sample sequence 
$\kappa^{\sample}_A:=(b^{(j_i)}_A\oplus b^{(j_i+1)}_A)_{i\in\mathcal{D}_{\sample}}$. 
\item
\label{BEC}
(Bit error correction) 
Alice estimates the bit error rate in the code rounds using the information of the one in the sample rounds. 
Depending on the estimated error rate, 
Alice chooses and announces a bit error correcting code and 
sends syndrome information on her sifted key $\kappa_A$ by consuming a pre-shared secret key of length $N_{\ec}$. 
Bob corrects the bit errors in his sifted key $\kappa_B$ and obtains the reconciled key $\kappa_B^{\rec}$. 
By consuming a pre-shared secret key of length $\zeta'$, Alice and Bob verify the correctness of their resulting reconciled keys 
by comparing the output ($\zeta'$-bit) of a randomly chosen universal$_2$ hash function $H_{\ec}$. 
\item
\label{sPA}
(Privacy amplification)
Alice and Bob conduct privacy amplification by shortening $N_{\pa}$ bits to respectively share the final keys $k_A$ and $k_B$ of length 
\begin{align}
N_{\fin}=N_{\U{code}}-N_{\pa}.
\label{eq:finalkey}
\end{align}
\end{enumerate}
We define two parameters which will be used in our security proof in Sec.~\ref{sec:upphase}. 
We define the detection rate by
\begin{align}
0\le Q:=\frac{N_{\det}}{N_{\emm}}\le1,
\label{def:Q}
\end{align}
and the bit error rate in the sample rounds  by
\begin{align}
e_{\bit}:=\frac{\wt(\kappa_A^{\sample}\oplus\kappa_B^{\sample})}{N_{\sample}}.
\label{biter}
\end{align}
The net length of the final key, namely, the increased length of the secret key is written as
\begin{align}
\ell=N_{\fin}-N_{\ec}-\zeta'.
\label{eq:Lkey}
\end{align}

\section{security proof}
\label{sec:secu}
In this section, we prove the security of the actual protocol described in Sec.~\ref{pro} in the finite-size regime. 
In Sec.~\ref{secA}, we explain in what sense we claim that the protocol is secure. 
Here, we adopt the universal composable security criterion~\cite{composable2009Renner}, 
which is widely used in the security proofs of QKD. 
In Sec.~\ref{sec:nnn}, we prove the security of the actual protocol based on the complementarity argument~\cite{koashi2009}.  
This argument reduces the security proof to estimating how well Alice can predict the outcome of the complementary observable, 
which is quantified by the number of phase errors. 
In Sec.~\ref{sec:upphase},  we estimate the upper bound on the number of phase errors where we leave its detailed statistical analysis to Appendix~\ref{app:rigo}. 
Note that another security proof framework based on the entropic uncertainty principle~\cite{eup} and 
the leftover hashing lemma~\cite{lhl} also reduces the proof to estimating the upper bound on the number of phase errors~\cite{AL,ncomm12}. 
Hence, using the discussions in Sec.~\ref{sec:upphase}, we can also prove the security within this framework.

We summarize the definitions used in this section. 
The projector is defined by $\hat{P}[\ket{x}]:=\ket{x}\bra{x}$, 
the 1-norm $||\hat{A}||_1$ for linear operator $\hat{A}$ by 
$||\hat{A}||_1:=\tr\sqrt{\hat{A}^{\dagger}\hat{A}}$, and function $h(x)$ by
\begin{align}
  h(x):=
    \begin{cases}
-x\log_2x-(1-x)\log_2(1-x) & (0\le x\le 1/2) \\
1 & (x>1/2).
  \end{cases}
  \label{binaryentropyfunction}
\end{align}
The Kronecker delta is defined by $\delta_{x,y}=1$ if $x=y$ and $\delta_{x,y}=0$ otherwise. 

\subsection{Security criterion}
\label{secA}
Here, we explain our security criterion that satisfies the universal composability~\cite{composable2009Renner}. 
When the length of the final key is $N_{\fin}$, we denote the state of Alice's and Bob's final keys and Eve's quantum system by
\begin{align}
&\hat{\rho}_{ABE|N_{\fin}}^{\rm fin}:=
\sum_{\substack{k_A,k_B\in\\\{0,1\}^{N_{\fin}}}}\Pr[k_A,k_B|N_{\fin}]|k_A,k_B\rangle\langle k_A,k_B|_{AB}\no\\
&\otimes\hat{\rho}^{\fin}_{E|N_{\fin}}(k_A,k_B), 
\label{eq:act}
\end{align}
and denote the one of the ideal final keys and Eve's quantum system by
\begin{align}
&\hat{\rho}_{ABE|N_{\fin}}^{\rm ideal}:=\no\\
&\frac{1}{2^{N_{\fin}}}\sum_{k\in\{0,1\}^{N_{\fin}}}|k,k\rangle\langle k,k|_{AB}\otimes
\tr_{AB}(\hat{\rho}^{\fin}_{ABE|N_{\fin}}).
\label{eq:ideal}
\end{align}
We say a protocol is $\epsilon_{\U{sec}}$-secure if 
\begin{align}
\frac{1}{2}\sum_{N_{\fin}\ge0}\Pr[N_{\fin}]||\hat{\rho}_{ABE|N_{\fin}}^{\rm ideal}-\hat{\rho}_{ABE|N_{\fin}}^{\rm fin}||_1\le\epsilon_{\U{sec}}.
\label{kokok}
\end{align}
Here, $\Pr[N_{\fin}]$ denotes the probability of obtaining the final key of length $N_{\fin}$ by executing the protocol, 
where aborting the protocol corresponds to $N_{\fin}=0$. 
We say the protocol is $\epsilon_c$-correct if 
\begin{align}
\sum_{N_{\fin}\ge0}\Pr[N_{\fin}]\Pr[k_A\neq k_B|N_{\fin}]\le\epsilon_c.
\label{eq:correctcond}
\end{align}
Also, we say the protocol is $\epsilon_s$-secret if 
\begin{align}
\frac{1}{2}\sum_{N_{\fin}\ge0}\Pr[N_{\fin}]||\hat{\rho}^{\fin}_{AE|N_{\fin}}-\hat{\rho}^{\U{ideal}}_{AE|N_{\fin}}||_1\le\epsilon_s.
\label{eq:secrecycond}
\end{align}
Here, we define 
\begin{align}
\hat{\rho}^{\fin}_{AE|N_{\fin}}:=\tr_B[\hat{\rho}^{\fin}_{ABE|N_{\fin}}]
\label{d1}
\end{align}
and 
\begin{align}
\hat{\rho}^{\U{ideal}}_{AE|N_{\fin}}:=\tr_B[\hat{\rho}^{\U{ideal}}_{ABE|N_{\fin}}]. 
\label{d2}
\end{align}
As shown in~\cite{koashi2009}, if the protocol is $\epsilon_c$-correct and $\epsilon_s$-secret, it is 
$\epsilon_{\sec}$-secure with
\begin{align}
\epsilon_{\sec}=\epsilon_c+\epsilon_s.
\label{eq:comb}
\end{align}
For completeness of this paper, we give the proof of Eq.~(\ref{eq:comb}) in Appendix~\ref{app:secrecyconddecomp}. 

As for correctness, due to verification of error correction executed in step~\ref{BEC}, 
the probability of obtaining different final keys is upper-bounded by $2^{-\zeta'}$~\cite{AL}. 
We state this as the following theorem, whose proof is given in Appendix~\ref{app:V}. 
\begin{theorem}
\label{th:correct}
(Correctness) 
The actual protocol described in section~\ref{pro} is $\epsilon_c$-correct with $\epsilon_c=2^{-\zeta'}$. 
\end{theorem}
In the following Secs.~\ref{sec:nnn} and \ref{sec:upphase}, our purpose is to derive the secrecy parameter defined in Eq.~(\ref{eq:secrecycond}). 

\subsection{Derivation of secrecy parameter}
\label{sec:nnn}
Here, we derive the upper bound on the secrecy parameter $\epsilon_s$ in Eq.~(\ref{eq:secrecycond}). 
In so doing, we consider virtual procedures equivalent to Alice's state preparation in step~\ref{pro:A}, 
the calculation of her sifted key $\kappa_A$ and sample sequence $\kappa_A^{\sample}$ in step~\ref{pr:bb} 
and Bob's measurements. These procedures simplify the derivation of $\epsilon_s$, and 
the final state $\hat{\rho}^{\fin}_{AE|N_{\fin}}$ of Alice's and Eve's systems is the same as the one of the actual protocol. 
As can be seen from Eq.~(\ref{eq:secrecycond}), Bob's system does not appear in the definition of the $\epsilon_s$-secret. 
Hence, we can consider that Bob virtually executes an operation such that it makes it easier to prove Eq.~(\ref{eq:secrecycond}). 
These virtual procedures are the same as those in our previous work~\cite{npjmizu2018}, and we concisely state them below.
 
As for the virtual procedure equivalent to step~\ref{pro:A}, Alice prepares three auxiliary qubits of systems $\bm{A}:=A_1A_2A_3$, generates state 
\begin{align}
\ket{\Phi}_{\bm{ASR}}:=2^{-3/2}\bigotimes_{u=1}^3\sum_{b^{(u)}_A=0,1}\hat{H}\ket{b^{(u)}_A}_{A_u}\ket{\psi_{b_A^{(u)}}}_{S_uR_u}
\end{align}
and sends system $\bm{S}$ to Bob. 
Here, $\hat{H}:=1/\sqrt{2}\sum_{x,y=0,1}(-1)^{xy}\ket{x}\bra{y}$, $\bm{R}:=R_1R_2R_3$, 
and $\ket{\psi_{b_A^{(u)}}}_{S_uR_u}$ is a purification of $\hat{\rho}^{b^{(u)}_A}_{S_u}$.

Regarding the virtual procedure for step~\ref{pr:bb}, Alice calculates bit $b^{(j)}_A\oplus b^{(j+1)}_A$ by applying controlled-not 
(CNOT) gate $\hat{U}^{(j)}_{\cnot}$ with $\hat{U}^{(j)}_{\cnot}\ket{x}_{A_j}\ket{y}_{A_{j+1}}:=\ket{x}_{A_j}\ket{x\oplus y}_{A_{j+1}}$ for 
$x,y\in\{0,1\}$ followed by measuring system $A_j$ in the $X$ basis. 
Here, we define $Z$- and $X$-basis states as $\{\ket{0},\ket{1}\}$ and $\{\ket{+},\ket{-}\}$ with $\ket{\pm}:=(\ket{0}\pm\ket{1})/\sqrt{2}$, respectively. 
 
In the complementary argument~\cite{koashi2009}, 
we are interested in how well Alice can predict the outcome 
$z_j\in\{0,1\}$ if system $A_j$ were measured in the $Z$ basis, which 
is the complementary basis of the key generation basis (namely, the $X$ basis). 
Here, we define $z_j$ as the $Z$-basis measurement outcome of system $A_j$ before performing $\hat{U}^{(j)}_{\cnot}$. 
As for $z_j$, since $\hat{U}^{(j)}_{\cnot}$ and $Z$-basis measurement of system $A_j$ commute, 
$z_j$ is regarded as the outcome of the same measurement after performing $\hat{U}^{(j)}_{\cnot}$. 
Bob's role is to help Alice's prediction of $z_j$. In particular, instead of Bob learning the key bit by interfering with the $j^{\U{th}}$ and $(j+1)^{\U{th}}$ pulses, 
he measures which of the two pulses contains a single photon, whose information is sent to Alice. 
Also, to predict $z_j$, Alice measures her system $A_{j+1}$ in the $Z$-basis after performing $\hat{U}^{(j)}_{\cnot}$. 
This gives Alice the information of the outcome $z_j\oplus z_{j+1}$. 
Note that this prediction strategy using Alice's and Bob's information is the same as the one of our previous analysis~\cite{npjmizu2018}. 
We define the occurrence of a {\it phase error} if her prediction fails. 
More precisely, Alice's task is to predict the $Z$-basis measurement outcomes $\bm{z}_{\code}:=(z_{j_i})_{i\in\mathcal{D}_{\code}}$ by 
using the information sent by Bob and Alice's information of $z_j\oplus z_{j+1}$ 
when the auxiliary qubits of systems $\bm{A}_{\code}:=(A_{j_i })_{i\in\mathcal{D}_{\code}}$ were measured in the $Z$ basis 
just after Bob completes all the detections. 
We denote the prediction of $\bm{z}_{\code}$ by $\bm{z}^*_{\code}$. 
Then, the complementarity argument~\cite{koashi2009,matsuura2019} claims that if 
$\bm{z}_{\code}\oplus \bm{z}^*_{\code}$ is in a set 
$\mathcal{T}_{\ph}\subset\{0,1\}^{N_{\code}}$ with unit probability, by shortening the reconciled key by
\begin{align}
N_{\pa}=\log_2|\mathcal{T}_{\ph}|+\zeta
\end{align}
for $\zeta>0$ in the privacy amplification step~\ref{sPA}, we obtain $\epsilon_s=\sqrt{2}\sqrt{2^{-\zeta}}$. 
If $\bm{z}_{\code}\oplus \bm{z}^*_{\code}$ is not in $\mathcal{T}_{\ph}$ with probability $\epsilon$, namely, 
\begin{align}
\Pr[\bm{z}_{\code}\oplus \bm{z}^*_{\code}\notin\mathcal{T}_{\ph}]\le\epsilon,
\label{eq:ZZ}
\end{align}
we have $\epsilon_s=\sqrt{2}\sqrt{\epsilon+2^{-\zeta}}$~\cite{koashi2009,matsuura2019}. 
One way to obtain Eq.~(\ref{eq:ZZ}) is to estimate the upper bound on the number $N_{\ph}$ of phase errors. 
That is, if we have
\begin{align}
\Pr[N_{\ph}>N^{\U{U}}_{\ph}]\le\epsilon
\end{align}
with $N^{\U{U}}_{\ph}$ being a function of experimentally available data, we obtain Eq.~(\ref{eq:ZZ}). 
This is simply because $N_{\ph}\le N^{\U{U}}_{\ph}$ leads to $\bm{z}_{\code}\oplus \bm{z}^*_{\code}\in\mathcal{T}_{\ph}$ by setting 
$\mathcal{T}_{\ph}=\{\bm{b}\in\{0,1\}^{N_{\code}}|\wt(\bm{b})\le N_{\ph}^{\U{U}}\}$, 
whose number of elements $|\mathcal{T}_{\ph}|$ is upper-bounded by $2^{N_{\code}h(N_{\ph}^{\U{U}}/N_{\code})}$ 
with $h(x)$ defined in Eq.~(\ref{binaryentropyfunction}). 
We summarize the arguments in this section as the following theorem. 

\begin{theorem}
\label{th:secrecy}
(Secrecy) 
For the protocol described in section~\ref{pro}, if the number of phase errors $N_{\ph}$ satisfies the following 
regardless of Eve's attack: 
\begin{align}
\Pr[N_{\ph}>N^{\U{U}}_{\ph}]\le\epsilon
\label{eq:upboundPH}
\end{align}
for $\epsilon$ ($0\le\epsilon\le1$) and $N^{\U{U}}_{\ph}$ being a function of experimentally available data, 
and if the amount of privacy amplification $N_{\pa}$ is set to be
\begin{align}
N_{\pa}=N_{\U{code}}h\left(\frac{N^{\U{U}}_{\ph}}{N_{\U{code}}}\right)+\zeta
\end{align}
for $\zeta>0$, the protocol is $\epsilon_s$-secret with  
\begin{align}
\epsilon_s=\sqrt{2}\sqrt{\epsilon+2^{-\zeta}}. 
\label{eq:epsilonsex}
\end{align}
\end{theorem}
By combining Eqs.~(\ref{eq:Lkey}) and (\ref{eq:comb}), and Theorems~\ref{th:correct} and \ref{th:secrecy}, we obtain the following corollary. 
\begin{coro}
\label{coro:keyrate}
For any $\zeta,\zeta'>0$ and 
under Eq.~(\ref{eq:upboundPH}), the protocol described in Sec.~\ref{pro} generates the secret key of length 
\begin{align}
\ell=N_{\U{code}}\left[1-h\left(\frac{N^{\U{U}}_{\ph}}{N_{\U{code}}}\right)\right]-\zeta-N_{\U{EC}}-\zeta'
\end{align}
with 
$\epsilon_{\sec}=2^{-\zeta'}+\sqrt{2}\sqrt{\epsilon+2^{-\zeta}}$-secure. 
\end{coro} 
To complete our security proof, the remaining task is to derive the upper bound $N^{\U{U}}_{\ph}$ 
as well as the failure probability $\epsilon$ of the estimation in Eq.~(\ref{eq:upboundPH}). 
Note that $N^{\U{U}}_{\ph}$ is a function of the parameter $q_n$, which characterizes the source, 
as well as of random variables, such as $N_{\emm}, N_{\det}, N_{\code}, N_{\sample}$, and $e_{\bit}$, 
all of which are actually observed in the experiment.

\subsection{Estimation of the number of phase errors and its failure probability}
\label{sec:upphase}
In this section, we derive the upper bound on the number of phase errors and the failure probability of its estimation. 
The result in this section is an extension of our previous Theorem~1 in~\cite{npjmizu2018} to the finite-size regime. 

We aim to estimate the number of phase errors in the code rounds using experimentally observed numbers. 
In so doing, we define POVM (positive operator valued measure) elements for obtaining the phase error event in the code round 
and the bit error event in the sample round. 
To define these POVMs, we introduce the POVM element for Bob's detected event. 
Given Bob obtains the detected event, POVM elements $\{\hat{\Pi}_{j,d}\}_{j,d}$ for detecting bit $d\in\{0,1\}$ 
at the $j^{\U{th}}$ ($j\in\{1,2\}$) time slot can be written as~\cite{npjmizu2018}
\begin{align}
\hat{\Pi}_{j,d}:=\hat{P}[\ket{\hat{\Pi}_{j,d}}_B]
\end{align}
with
\begin{align}
\ket{\hat{\Pi}_{j,d}}_B:=\frac{\sqrt{w_j}\ket{j}_B+(-1)^d\sqrt{w_{j+1}}\ket{j+1}_B}{\sqrt{2}},
\end{align}
where $w_1=w_3=1$ and $w_2=1/2$. 
Here, $\{\ket{i}_B\}_{i=1}^3$ denotes the orthogonal states, where $\ket{2}_B$ represents that the 2$^{\U{nd}}$ incoming pulse has a single photon, and 
$\ket{1}_B$ ($\ket{3}_B$) represents that the $1^{\U{st}}$ pulse passing the long arm 
($3^{\U{rd}}$ pulse passing the short arm) of the first BS in the Mach-Zehnder interferometer contains a single photon. 
As explained in Sec.~\ref{sec:nnn}, the phase error event occurs when Alice fails the prediction of the $Z$-basis measurement outcome $z_j$ of system $A_j$. 
The explicit formula of POVM element $\hat{e}_{\ph}$ corresponding to obtaining 
the phase error event is the same as our previous work~\cite{npjmizu2018}, which is given by
\begin{align}
&\hat{e}_{\ph}=\sum_{j=1}^2\sum_{\bm{z}}\hat{P}[\ket{\bm{z}}_{\bm{A}}]\otimes\no\\
&\left[w_j\delta_{z_{j+1},1}\hat{P}[\ket{j}_B]+w_{j+1}\delta_{z_{j},1}
\hat{P}[\ket{j+1}_B]\right]
\label{eqPh}
\end{align}
with $\bm{z}:=z_1z_2z_3\in\{0,1\}^3$. 
Since $\hat{e}_{\ph}$ is diagonal in the basis $\ket{\bm{z}}_{\bm{A}}$, the measurement of the weight $a:=\wt(\bm{z})$, namely, 
$\{\hat{P}_a\}_{a\in\{0,1,2,3\}}$ with
\begin{align}
\hat{P}_a:=\sum_{\bm{z}:\wt(\bm{z})=a}\hat{P}[\ket{\bm{z}}_{\bm{A}}]
\label{eqPA}
\end{align}
and $\{\hat{e}_{\ph},I_{AB}-\hat{e}_{\ph}\}$ commute. 
To relate the probability of obtaining a phase error with the one of a bit error, we also introduce the POVM element $\hat{e}_{\bit}$ 
corresponding to obtaining a bit error. This is given by~\cite{npjmizu2018}
\begin{align}
&\hat{e}_{\bit}=\sum_{j=1}^2\Bigg[\left(\hat{P}[\ket{++}_{A_jA_{j+1}}]+\hat{P}[\ket{--}_{A_jA_{j+1}}]\right)\otimes\hat{\Pi}_{j,1}\no\\
&+\left(\hat{P}[\ket{+-}_{A_jA_{j+1}}]+\hat{P}[\ket{-+}_{A_jA_{j+1}}]\right)\otimes\hat{\Pi}_{j,0}
\Bigg].
\label{eqPb}
\end{align}
Then, thanks to Lemmas~1 and 2 in~\cite{npjmizu2018}, we have the relation between the probabilities of 
obtaining a phase error, a bit error and the weight $a$ as 
\begin{align}
\tr[\hat{e}_{\ph}\hat{\sigma}_{AB}]&\le\lambda
\left(\tr[\hat{e}_{\bit}\hat{\sigma}_{AB}]+\sqrt{\tr[\hat{\sigma}_{AB}\hat{P}_1]
\cdot\tr[\hat{\sigma}_{AB}\hat{P}_3]}\right)\no\\
&+\sum_{a=2,3}\tr[\hat{\sigma}_{AB}\hat{P}_a]
\label{eq:relationpro}
\end{align}
with $\lambda:=3+\sqrt{5}$. 
Importantly, this inequality holds for any state $\hat{\sigma}_{AB}$ of systems $AB$. 

\begin{figure}[t]
\includegraphics[width=8cm]{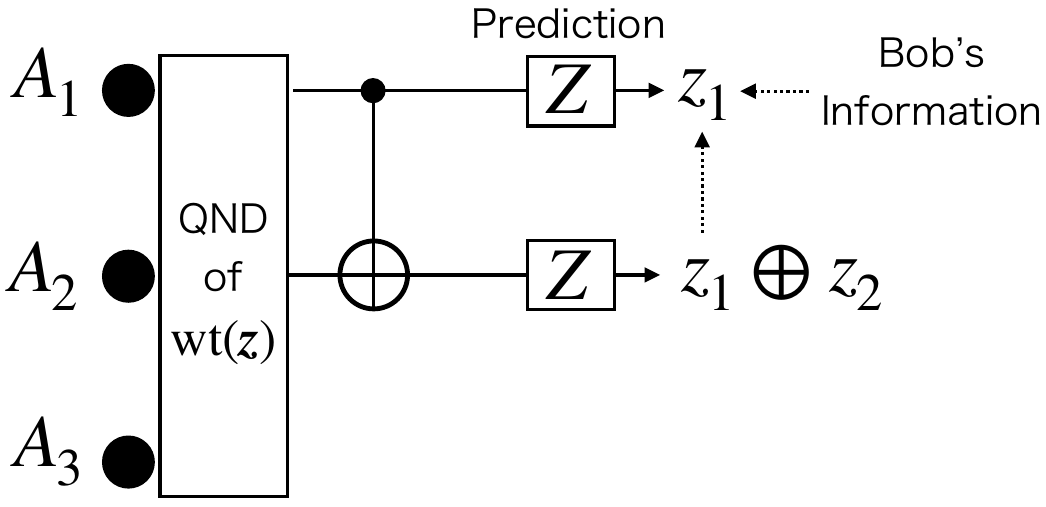}
\caption{
Schematics of Alice's measurement procedures for $j=1$ when $\chi^{(i)}_{\U{r}}=0$, i.e., the code rounds. 
She first carries out the QND measurement to learn the weight $\U{wt}(\bm{z})$ followed by performing the CNOT gate, and 
using the information of $z_1\oplus z_2$ and the one sent by Bob, Alice predicts the outcome $z_1$. 
Recall that the phase error event is the one if this prediction fails. }
 \label{virtualAlice}
\end{figure}

\begin{figure}[t]
\includegraphics[width=8.8cm]{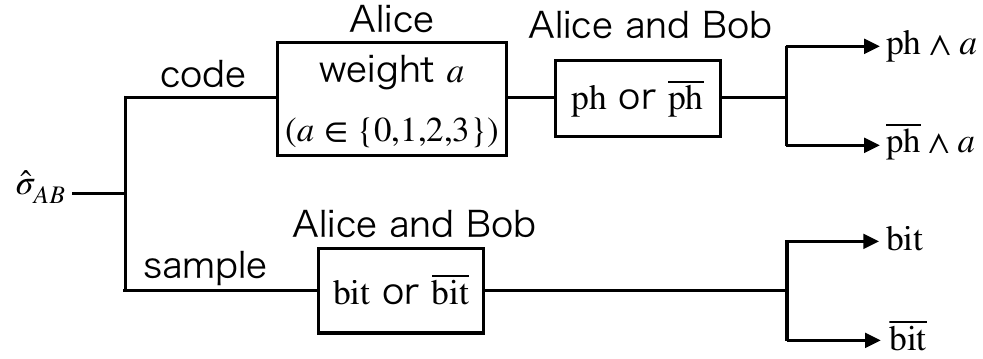}
\caption{
This figure illustrates the measurements performed by Alice and Bob in the code and sample rounds, respectively. 
}
 \label{diagram}
\end{figure}

To derive the upper bound on the number of phase errors $N_{\ph}^{\U{U}}$, we consider the following stochastic trial of 
measuring $N_{\det}$ systems $AB$. 
As described in step~\ref{tt} of the actual protocol, 
Bob probabilistically associates each detected round with the code or sample one with probability $t$ or $1-t$, respectively. 
For the code rounds, Alice and Bob extract the secret key while the sample ones are used to learn bit error rate $e_{\bit}$ defined in Eq.~(\ref{biter}). 
We define the random variable $\chi^{(i)}_{\U{r}}\in\{0,1\}$, 
which takes the value of 0 (1) if the $i^{\U{th}}$ detected round with $i\in\mathcal{D}$ is the code (sample) one. 
If $\chi^{(i)}_{\U{r}}=0$, Alice 
carries out the quantum nondemolition (QND) measurement on her three qubits associated with the $i^{\U{th}}$ detected round to learn the weight 
$\wt(\bm{z})$ with the POVM $\{\hat{P}_a\}_{a=0}^3$ defined in Eq.~(\ref{eqPA}). 
After that Alice and Bob measure their systems to know whether the $i^{\U{th}}$ detected round has a phase error or not by 
the POVM $\{\hat{e}_{\ph},\hat{I}_{AB}-\hat{e}_{\ph}\}$ defined in Eq.~(\ref{eqPh}). 
Recall that such simultaneous measurements are allowed because $\hat{e}_{\ph}$ and $\hat{P}_a$ commute for any $a\in\{0,1,2,3\}$. 
Fig.~\ref{virtualAlice} depicts Alice's measurement procedures when $\chi^{(i)}_{\U{r}}=0$. 
On the other hand, if 
$\chi^{(i)}_{\U{r}}=1$, Alice and Bob measure their systems to know whether the $i^{\U{th}}$ detected round has a bit error or not 
by the POVM $\{\hat{e}_{\bit},\hat{I}_{AB}-\hat{e}_{\bit}\}$ defined in Eq.~(\ref{eqPb}). 
These situations are illustrated in Fig.~\ref{diagram}. 

In this stochastic trial, the $i^{\U{th}}$ measurement outcome is in set 
\begin{align}
\mathcal{S}:=\bigcup_{a=0}^3\{\ph\wedge a,\overline{\ph}\wedge a\}\cup\{\bit,\overline{\bit}\}, 
\end{align}
where ``ph" (``bit") denotes the measurment outcome that the $i^{\U{th}}$ detected round entails the phase (bit) error, and 
``$\overline{\ph}$" (``$\overline{\bit}$") has no-phase (no-bit) error. 
We introduce the following random variables $\chi^{(i)}_{\U{ph}},\{\chi^{(i)}_{a}\}_{a=0}^3$ and $\chi^{(i)}_{\U{bit}}$, 
each of which takes the value of 0 or 1 according to the $i^{\U{th}}$ measurement outcome:
\begin{align}
\chi^{(i)}_{\U{ph}}= 
\begin{cases} 1  & \mbox{if the}~i^{\U{th}}~\mbox{measurement outcome is ph},\\
0 & \mbox{otherwise},
\end{cases}
\end{align}
\begin{align}
\chi^{(i)}_{a}= 
\begin{cases} 1  & \mbox{if the}~i^{\U{th}}~\mbox{measurement outcome is}~a,\\
0 & \mbox{otherwise},
\end{cases}
\end{align}
and
\begin{align}
\chi^{(i)}_{\U{bit}}= 
\begin{cases} 1  & \mbox{if the}~i^{\U{th}}~\mbox{measurement outcome is bit},\\
0 & \mbox{otherwise}.
\end{cases}
\end{align}
We also introduce $\{F^{(i)}\}_{i=0}^{N_{\det}}$ as the filtration with $F^{(i)}$ identifying the random variables including 
$\chi^{(i')}_{\U{ph}}, \{\chi^{(i')}_{a}\}_{a=0}^3$ and $\chi^{(i')}_{\U{bit}}$ for $i'\in\{1,2,...,i\}$. 
That is, $F^{(i)}\subseteq 2^{\Omega}$ is a $\sigma$-algebra on sample space 
$\Omega=\mathcal{S}^{\times N_{\det}}$, which satisfies 
$F^{(i)}\subseteq F^{(j)}$ and $E[X^{(i)}|F^{(j)}]=X^{(i)}$ for $i\le j$ and a sequence of random variables $\{X^{(i)}\}_i$
~\cite{footnote1}. 
Although the elements of $F^{(j)}$ are events, identifying one element of $F^{(j)}$ is equivalent to 
identifying the first $j$ measurement outcomes. 
Therefore, $E[X^{(i)}|F^{(j)}]$ is regarded as the expectation of $X^{(i)}$ conditioned on the first $j$ measurement outcomes. 
Then, the conditional expectations of random variables $\chi^{(i)}_{\U{ph}}, \chi^{(i)}_{a}$ and $\chi^{(i)}_{\U{bit}}$ are respectively given by
\begin{align}
E[\chi^{(i)}_{\U{ph}}|F^{(i-1)}]&=t\cdot\tr[\hat{e}_{\ph}\hat{\sigma}^{F^{(i-1)}}_{AB}],
\label{eq:ig1}
\\
E[\chi^{(i)}_{a}|F^{(i-1)}]&=t\cdot\tr[\hat{P}_{a}\hat{\sigma}^{F^{(i-1)}}_{AB}], 
\label{eq:ig2}
\\
E[\chi^{(i)}_{\U{bit}}|F^{(i-1)}]&=(1-t)\cdot\tr[\hat{e}_{\bit}\hat{\sigma}^{F^{(i-1)}}_{AB}]. 
\label{eq:ig3}
\end{align}
Here, $\hat{\sigma}^{F^{(i-1)}}_{AB}$ denotes the state of systems $AB$ conditional on the first $(i-1)$ measurement outcomes. 
Since Eq.~(\ref{eq:relationpro}) holds for any state $\hat{\sigma}_{AB}$, Eq.~(\ref{eq:relationpro}) can be rewritten by using 
the conditional expectations as
\begin{align}
&\frac{E[\chi^{(i)}_{\U{ph}}|F^{(i-1)}]}{t}-\frac{\lambda}{1-t}E[\chi^{(i)}_{\U{bit}}|F^{(i-1)}]-\sum_{a=2,3}\frac{E[\chi^{(i)}_a|F^{(i-1)}]}{t}\no\\
&\le\frac{\lambda}{t}\sqrt{E[\chi^{(i)}_1|F^{(i-1)}]\cdot E[\chi^{(i)}_3|F^{(i-1)}]}. 
\label{nkk}
\end{align}
Taking the sum of the conditional expectations over all the detected events and using the Cauchy-Shwartz inequality lead to
\begin{align}
&\sum_{i=1}^{N_{\det}}
\Bigg(\frac{1}{t}E[\chi^{(i)}_{\U{ph}}|F^{(i-1)}]-\frac{\lambda}{1-t}E[\chi^{(i)}_{\U{bit}}|F^{(i-1)}]\no\\
&-\frac{1}{t}\sum_{a=2,3}E[\chi^{(i)}_a|F^{(i-1)}]\Bigg)\no\\
&\le\frac{\lambda}{t}\sqrt{
\left(\sum_{i=1}^{N_{\det}}E[\chi^{(i)}_1|F^{(i-1)}]\right)
\left(\sum_{i=1}^{N_{\det}}E[\chi^{(i)}_3|F^{(i-1)}]\right)}.
\label{eq:procond}
\end{align}
Next, we transform this inequality into the one in terms of the random variables 
$\sum_{i=1}^{N_{\det}}\chi^{(i)}_{\U{ph}}, \sum_{i=1}^{N_{\det}}\chi^{(i)}_{\U{bit}}$, and $\sum_{i=1}^{N_{\det}}\chi^{(i)}_{a}$. 
In so doing, we exploit two concentration inequalities, Azuma's~\cite{Azuma} and Kato's inequalities~\cite{kato}, which can be applied 
to correlated random variables. Azuma's inequality is a typical technique to bound the sum of conditional expectations with 
the number of occurrences and is widely used in the security proofs of QKD~\cite{azuma0,azuma001,azuma01,azuma1,azuma12,azuma2}. 
The explicit statement of this inequality is shown in Appendix~\ref{sec:azuma}. 
The deviation term of Azuma's inequality scales with $\sqrt{N_{\det}}$, which 
is independent of the magnitude of the target sum of conditional expectations. 
Hence, if this target sum is comparable to the deviation term, Azuma's inequality gives a reasonably tight bound. 
Unfortunately, however, if this sum is much smaller than the number of trials, this inequality only provides a loose bound. 
In our analysis, this is the case when we bound the following sum of conditional expectations
\begin{align}
S_{3}:=
\sum_{i=1}^{N_{\det}}E[\chi^{(i)}_3|F^{(i-1)}]
\label{nnkt}
\end{align}
with random variable $N_3:=\sum_{i=1}^{N_{\det}}\chi^{(i)}_3$. 
The number of the weight being three (namely, $a=3$) implies that the state of a single block contains at least three photons~\cite{footnote2}, 
whose probability of occurrence is much smaller than one. 
This means that $S_3$ is generally much smaller than $N_{\det}$, and hence Azuma's inequality only provides a loose bound on $S_3$. 

On the other hand, Kato's inequality is the recently found novel concentration inequality that always gives a tighter bound than Azuma's inequality and 
is employed in recent finite-key analyses~\cite{npj21,hyong,guPRA,ACqst}. 
Kato's inequality has a significant advantage over Azuma's one especially 
when the target sum of conditional expectations is much smaller than the number of trials. 
This advantage is brought by incorporating our prediction $N^\ast_{3}$ of $N_{3}$ into the estimation of $S_{3}$. 
The accuracy of this prediction only affects the tightness of this inequality, which 
is tightest when $N^\ast_{3}=N_{3}$, and the inequality is still valid even if the prediction fails. 
In our security analysis, we found that Kato's inequality indeed improves the key rate by tightly estimating $S_3$ 
for which Azuma's inequality is not tight. 
On the other hand, Azuma's inequality is sufficiently tight for all the other components whose conditional expectations are larger. 
For instance, in estimating the following sum of conditional expectations 
\begin{align}
S_2:=\sum_{i=1}^{N_{\det}}\sum_{a=2,3}E[\chi^{(i)}_a|F^{(i-1)}], 
\end{align}
Azuma's inequality gives a tight bound because $S_2$ is generally much larger than $S_3$. 
As a result, we confirm that the probability of multiple photon emission events, in which two or more than two photons are emitted, is not small enough 
to benefit significantly from Kato's inequality. 
This result implies that the finite-key analysis of the BB84 protocol with non-phase randomized light sources 
could not be drastically improved by applying Kato's inequality instead of Azuma's because the number of phase errors is given by 
the numbers of bit errors and essentially the multiple photon emission events~\cite{lopreskill}. 
On the other hand, the key rate of the twin-field protocol~\cite{npj21} is substantially increased by using Kato's inequality to estimate the sum of the expectations of vacuum detections. This is so because the vacuum detection occurs with about the dark count probability 
(such as $10^{-8}$ assumed in~\cite{npj21}), and this event is rare enough to benefit significantly from this inequality. 
To summarize the discussion so far, Kato's inequality could drastically improve the key rate of QKD protocols 
if the derived number of phase errors contains the number of occurrences of very rare events, 
such as vacuum detection and three-photon emission events.

In Appendix~\ref{sec:kato}, we explain how to apply Kato's inequality to bound $S_{3}$ with $N_3$ and its prediction $N_3^{\ast}$. 
We leave the details of deriving the upper bound on the number of phase errors $N_{\ph}^{\U{U}}$ 
from Eq.~(\ref{eq:procond}) to Appendix~\ref{app:rigo} and just state our main result as follows. 

\begin{widetext}
\begin{theorem} 
\label{th:mainPH}
For the protocol described in Sec.~\ref{pro}, the number of phase errors satisfies
\begin{align}
&N_\ph\le N^{\U{U}}_{\ph}:=
\frac{\lambda te_{\bit}N_{\sample}}{1-t}+\left(tq_2N_{\emm}+\Gamma_2\right)
+\lambda\times\no\\
&\sqrt{\left[tq_1N_{\emm}+\Gamma_1+\Delta(1,\epsilon_1)\right]
\left\{
\left(tq_3N_{\emm}+\Gamma_3\right)\left(1+\frac{2a^*(N_{\det},N_{3}^\ast,\epsilon_1)}{\sqrt{N_{\det}}}\right)+
[b^*(N_{\det},N_{3}^\ast,\epsilon_1)-a^*(N_{\det},N_{3}^\ast,\epsilon_1)]\sqrt{N_{\det}}
\right\}}\no\\
+&t\Delta(D^2,\epsilon_1)
\label{nge}
\end{align}
except for probability $3\epsilon_1+3\epsilon_2$ with $\epsilon_1$ and $\epsilon_2$ being the failure probabilities of the parameter estimation steps. 
Here, 
\begin{align}
\Delta(x,y)&:=\sqrt{2xN_{\det}\ln\frac{1}{y}},\\
\lambda&:=3+\sqrt{5},\\
\Gamma_n&:=\frac{1}{2}\left[-\ln\epsilon_2+\sqrt{\left(\ln\epsilon_2\right)^2-8tq_nN_{\U{em}}\ln\epsilon_2}\right],\\
N_3^\ast&=\min\left\{tq_3N_{\U{em}}+\Gamma_3,\lfloor (N_{\det}-1)/2 \rfloor\right\}, 
\\
D&:=\max\left\{\frac{\lambda}{1-t}+1,\frac{1}{t}+\lambda+1\right\},\\
a^\ast(n,m,\epsilon)&:=
\max\left\{-\frac{\sqrt{n}}{2},
\frac{216\sqrt{n}m(n-m)\ln\epsilon-48n^{\frac{3}{2}}(\ln\epsilon)^2
+27\sqrt{2}(n-2m)\sqrt{-n^2(\ln\epsilon)[9m(n-m)-2n\ln\epsilon]}}
{4(9n-8\ln\epsilon)[9m(n-m)-2n\ln\epsilon]
}\right\}
\label{ang}\\
b^\ast(n,m,\epsilon)&:=
\frac{\sqrt{18a^\ast(n,m,\epsilon)^2n-[16a^\ast(n,m,\epsilon)^2+24a^\ast(n,m,\epsilon)\sqrt{n}+9n]\ln\epsilon}}{3\sqrt{2n}}.
\label{bng}
\end{align}
Recall that probability $q_n$ is defined in Eq.~(\ref{eq:qa}). 
\end{theorem}
\end{widetext}
With this theorem, we complete the derivation of Eq.~(\ref{eq:upboundPH}) and our security proof.

\section{Simulations of key rates}
\label{sec:simu}
\begin{figure*}[t]
\includegraphics[width=11cm]{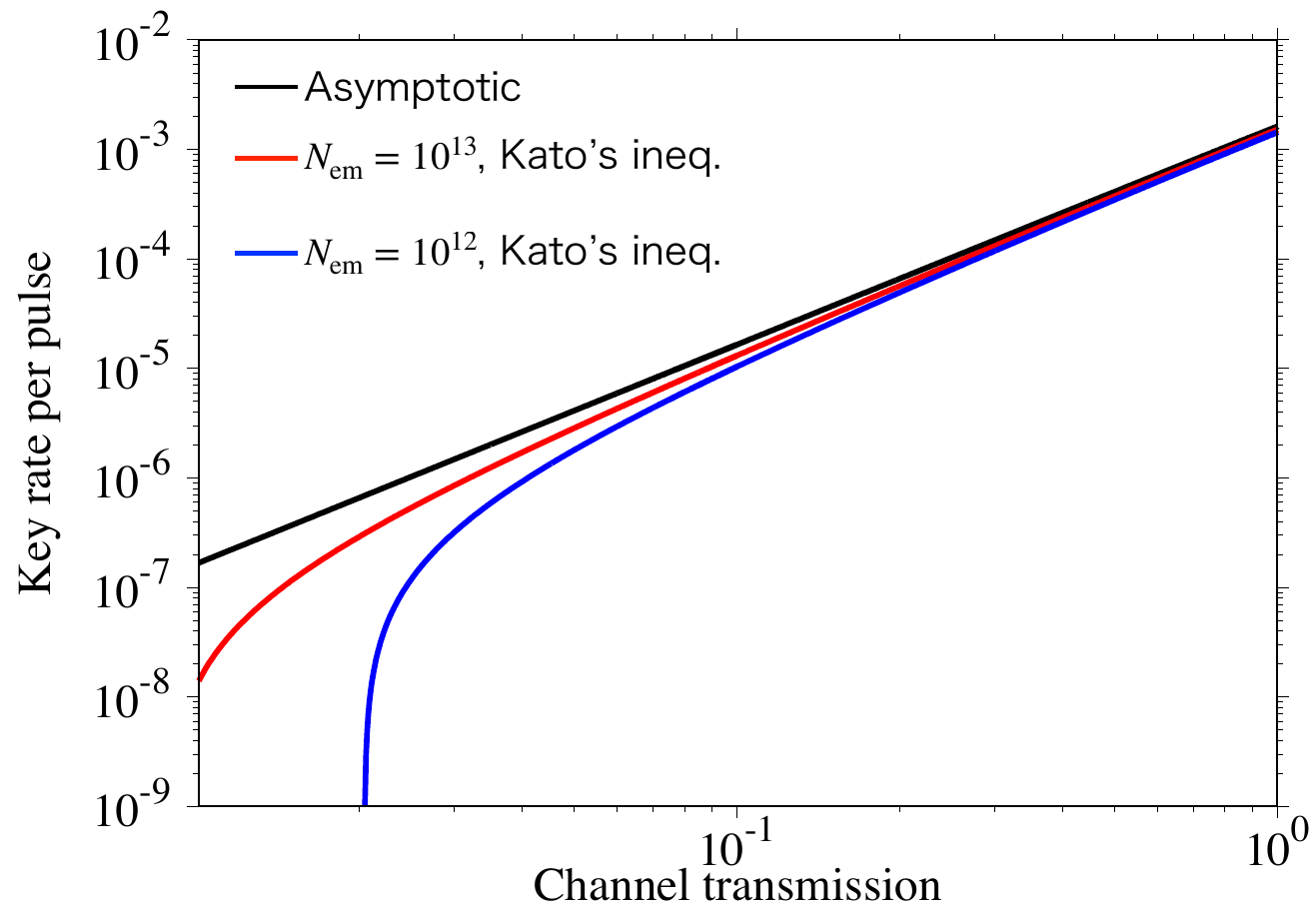}
\caption{
Secure key rate $R$ per a single emitted pulse as a function of the overall channel transmission $\eta$. 
From bottom to top, we plot the key rates for $N_{\U{em}}=10^{12}$, $10^{13}$, and the asymptotic case under the bit error rate of $e_{\bit}=1\%$ and 
the security parameter of $\epsilon_{\sec}\fallingdotseq 10^{-8.1}$. 
}
 \label{fig:key}
\end{figure*}
\begin{figure*}[t]
\includegraphics[width=11cm]{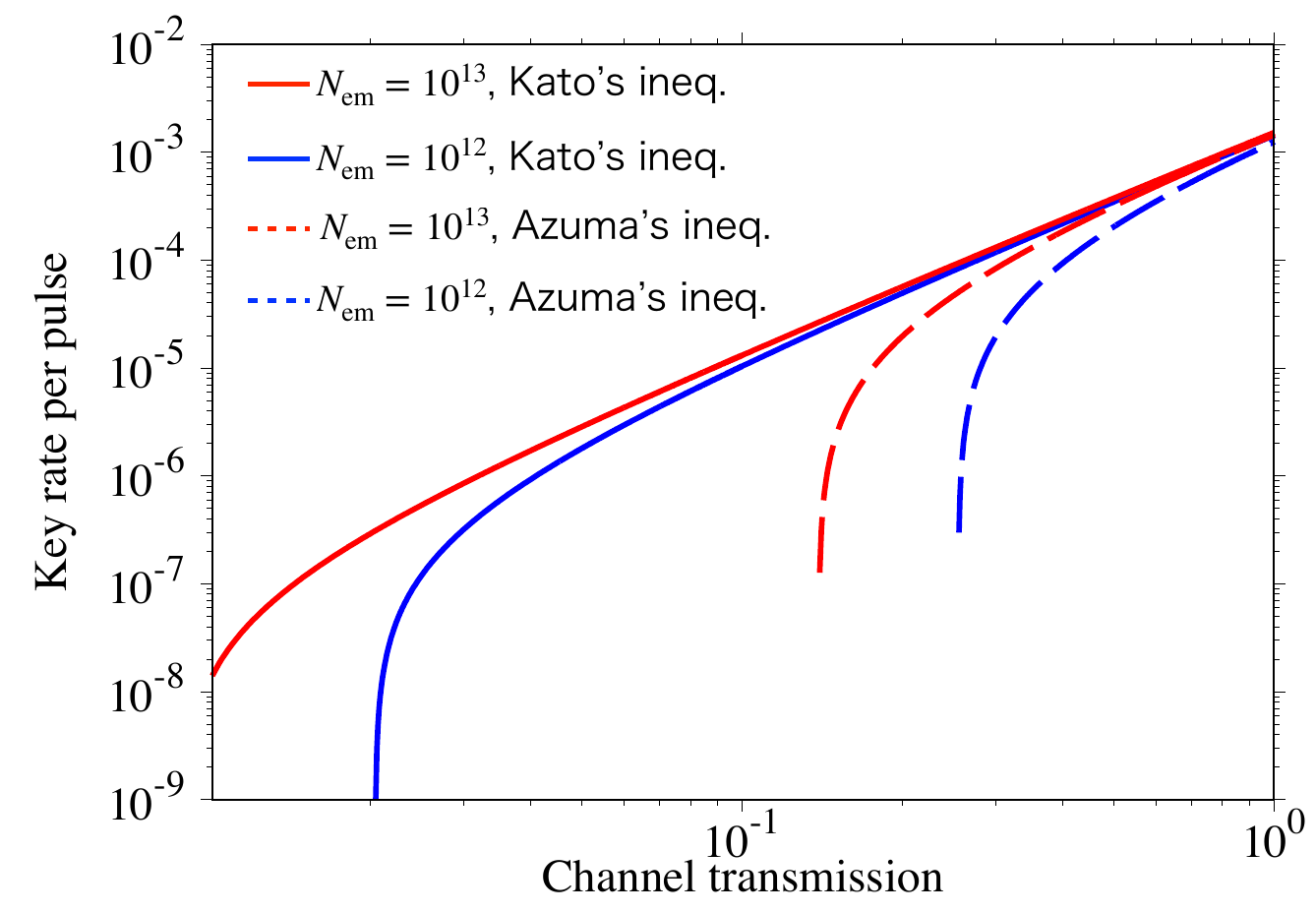}
\caption{
Secure key rate $R$ per a single emitted pulse as a function of the overall channel transmission $\eta$. 
The top two curves are the key rates for $N_{\U{em}}=10^{13}$ and $10^{12}$ under $e_{\bit}=1\%$ and $\epsilon_{\sec}\fallingdotseq 10^{-8.1}$ when 
Kato's inequality is used in the security proof 
(these two curves are the same as the two bottom curves in Fig.~\ref{fig:key}). 
The two bottom dashed curves in this figure are the key rates for $N_{\U{em}}=10^{12}$ and $10^{13}$ 
from bottom to top under $e_{\bit}=1\%$ and $\epsilon_{\sec}\fallingdotseq 10^{-8.1}$ when Azuma's inequality is used instead of Kato's one. 
We see that Azuma's inequality gives a slow convergence of the key rate in the finite-size regime, whose reason is 
discussed in Sec.~\ref{sec:ad}. 
}
 \label{fig:comparison}
\end{figure*}

\begin{figure*}[t]
\includegraphics[width=11cm]{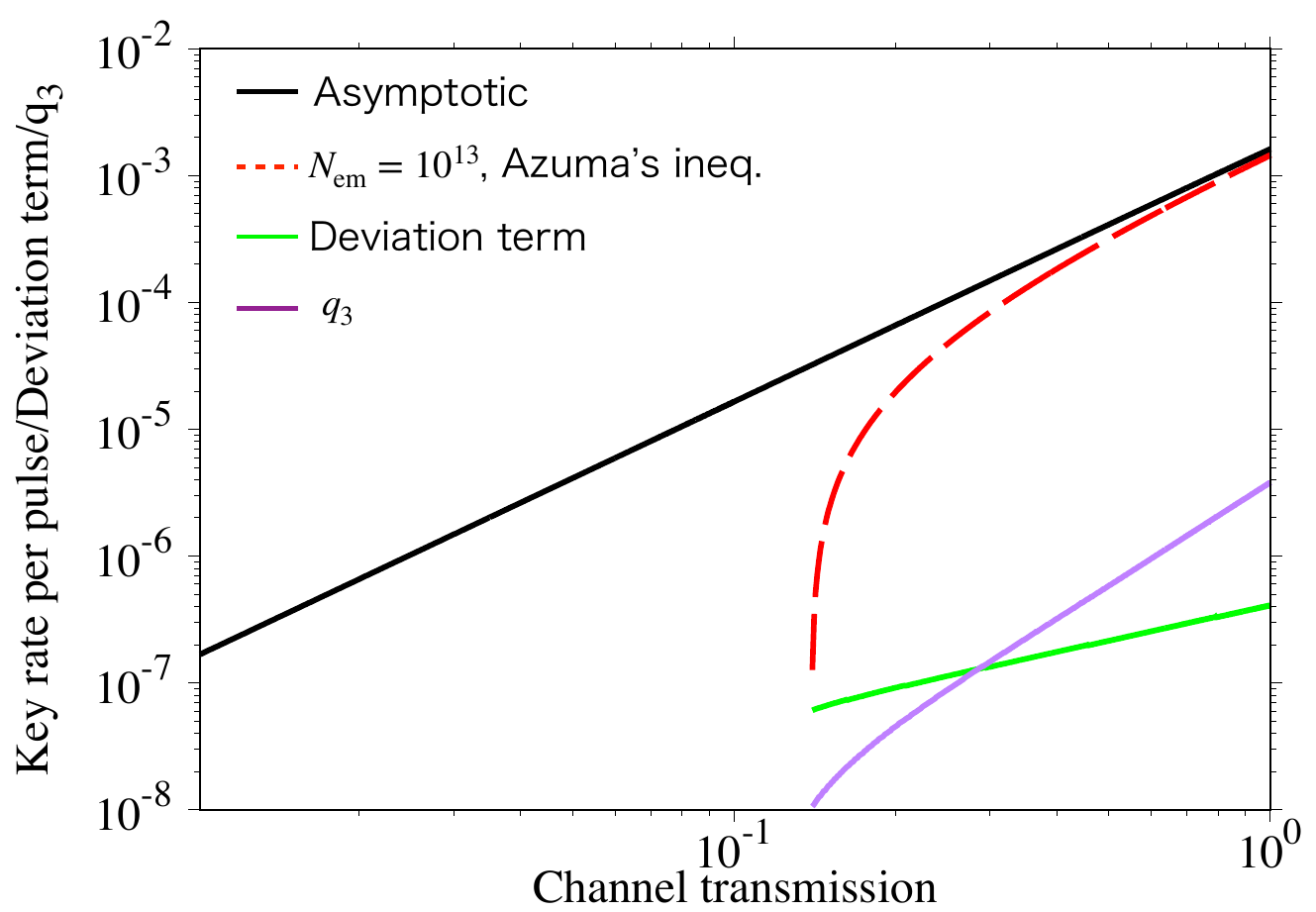}
\caption{
This figure illustrates why Azuma's inequality results in slow convergence of the key rate. 
The top black line is the asymptotic key rate (same as the black line in Fig.~\ref{fig:key}), and 
the second top curve is the one with $N_{\U{em}}=10^{13}$ when Azuma's inequality is used 
(same as the red dashed curve in Fig.~\ref{fig:comparison}). 
The bottom purple and green lines represent $q_3$ and the deviation term $\Delta(1,\epsilon_1)/tN_{\U{em}}$ of Azuma's inequality 
with $N_{\U{em}}=10^{13}$, respectively. 
The green and purple lines intersect at about $\eta=0.29$, and after this point 
the deviation term becomes dominant and the divergence of the two key rates increases drastically. 
}
 \label{fig:muD}
\end{figure*}
In this section, we present the simulation results of the key rate $R:=\ell/3N_{\U{em}}$ of our DPS protocol 
as a function of the channel transmission $\eta$ including the detection efficiency. 
From Corollary~\ref{coro:keyrate} and Theorem~\ref{th:mainPH}, $\ell$ can be expressed as
\begin{align}
\ell=N_{\U{code}}\left[1-h(N_{\ph}^{\U{U}}/N_{\code})\right]-\zeta-N_{\U{EC}}-\zeta'
\end{align}
with $\epsilon_{\sec}=2^{-\zeta'}+\sqrt{2}\sqrt{3\epsilon_1+3\epsilon_2+2^{-\zeta}}$-secure. 
For our simulation, we suppose that each emitted pulse is a coherent pulse from a laser with the mean photon number $\mu$. 
In this case, $q_a$ defined in Eq.~(\ref{eq:qa}) is written as 
\begin{align}
q_a=\sum_{\nu=a}^{\infty}e^{-3\mu}(3\mu)^{\nu}/\nu!. 
\label{eq:pnd}
\end{align}
We assume the number of detected rounds as 
\begin{align}
N_{\det}&=N_{\U{em}}\times 2\eta\mu e^{-2\eta\mu},\no\\
N_{\U{code}}&=tN_{\det},\no\\
N_{\U{samp}}&=(1-t)N_{\det},
\end{align}
and the practical cost of error correction being $N_{\ec}=1.16N_{\U{code}}h(e_\bit)$ with $1.16$~\cite{cascade}
is an error correction inefficiency. 
Also, we set $\zeta'=28$, $\zeta=58$, and $\epsilon_1=\epsilon_2=2^{-58}/6$, which results in 
$\epsilon_{\sec}=2^{-27}\fallingdotseq 10^{-8.1}$. 
The key rate $R$ is optimized over the mean photon number $\mu$ and the probability $t$ 
of choosing the code round in step~\ref{tt} for each value of $\eta$. 
The results are shown in Fig.~\ref{fig:key}. 
The optimal mean photon number $\mu_{\opt}$ against the channel transmission $\eta$ when $N_{\U{em}}=10^{13}$ is 
$(\eta,\mu_{\opt})=(1,9.3\times10^{-3}), (0.1,9.4\times10^{-4}),(0.01,9.0\times10^{-5})$. 
From the result with $N_{\U{em}}=10^{13}$ in Fig.~\ref{fig:key}, 
if we assume the overall channel transmission as $\eta=0.5\times 10^{-0.2l/10}$ with $l$ denoting the distance between Alice and Bob 
and laser diodes operating at 1 GHz repetition rate, by running our protocol for 8.3 hours, 
we can generate a 3 Mbit secret key for a channel length of 77 km under the bit error rate of 1\%.

\section{Comparisons of key rates with Azuma's and Kato's inequalities}
\label{sec:ad}
As explained in Sec.~\ref{sec:upphase}, we apply Kato's inequality to upper-bound $S_{3}$ in Eq.~(\ref{nnkt}). 
We remark that 
the security proof is valid even if we instead use Azuma's inequality to bound $S_3$, and in this case, the final expression of $N_{\ph}^{\U{U}}$ in 
Eq.~(\ref{nge}) is replaced with
\begin{align}
&N_{\ph}^{\U{U}}=
\frac{\lambda te_{\bit}N_{\sample}}{1-t}+\left(tq_2N_{\emm}+\Gamma_2\right)
+\lambda tN_{\emm}\times\no\\
&\sqrt{\left[q_1+\frac{\Gamma_1+\Delta(1,\epsilon_1)}{tN_{\emm}}\right]
\left[q_3+\frac{\Gamma_3+\Delta(1,\epsilon_1)}{tN_{\emm}}\right]}
+t\Delta(F^2,\epsilon_1).
\label{netuazuma}
\end{align}
To see how much the key rate degrades if we instead use this bound for the simulation of the key rate, 
we compare the key rate based on the bound in Eq.~(\ref{nge}) with the one based on Eq.~(\ref{netuazuma}) in Fig.~\ref{fig:comparison}. 
From this figure, it is clear that Kato's inequality gives a substantially better key rate in the finite-size regime. 
The slow convergence of the key rate using Azuma's inequality is due to the deviation 
term $\Delta(1,\epsilon_1)/tN_{\U{em}}$ in $q_{3}+\frac{\Gamma_3+\Delta(1,\epsilon_1)}{tN_{\U{em}}}$. 
Here, $q_3$ in Eq.~(\ref{eq:pnd}) is in the order of $O(\mu^3)\sim O(\eta^3)$, and 
the deviation term is in the order of 
\begin{align}
\frac{\Delta(1,\epsilon_1)}{tN_{\U{em}}}=\frac{1}{t}\sqrt{\frac{2\ln\frac{1}{\epsilon_1}}{N_{\U{em}}}}\sqrt{Q}
=O(\sqrt{Q})=O(\sqrt{\eta\mu})=O(\eta)
\end{align}
since $\mu=O(\eta)$. 
We illustrate in Fig.~\ref{fig:muD} the comparison of $q_3$ and $\Delta(1,\epsilon_1)/tN_{\U{em}}$, and the 
asymptotic key rate and the finite one using Azuma's inequality. 
From this figure, we see that the two lines of  $q_3\sim O(\eta^3)$ and $\Delta(1,\epsilon_1)/tN_{\U{em}}\sim O(\eta)$ 
intersect at $\eta \fallingdotseq0.29$, 
and the deviation term becomes dominant when $\eta<0.29$. This is the reason for drastically increasing the 
divergence of the two key rates after the point of $\eta=0.29$.

Note that from Ref.~\cite{kato}, 
when we increase the intensity and three photons are more likely to be emitted, the difference in the deviation terms of Kato's and Azuma's inequalities becomes smaller. Hence, one may expect that  the difference in the key rates under these two inequalities also becomes smaller by increasing the intensity. However, we do not observe such a tendency. 
This implies that the improvement we would obtain by increasing the 
intensities to decrease the deviations terms is overwhelmed by the use of non-optimal intensities.

\section{Conclusions}
\label{sec:conc}
\begin{table*}
 \begin{center}
       \caption{Comparison of our finite-key security analysis and the one in Ref.~\cite{DPSeacc}}
\begin{tabular}{ccccc} \hline
    & Eve's attack    &Key rate  per pulse & Sources &Detectors
    \\ \hline
   Our proof & Unconditional & $O(\eta^2)$ &Any source satisfying \ref{ASS1}-\ref{ASS3}
   &PNR detectors \\
   Ref.~\cite{DPSeacc}& Relativistic constraint& $O(\eta)$
  &Coherent states $\{\ket{\pm\alpha}\}$ &Threshold detectors
  \\ \hline
       \label{table:I}
\end{tabular}
\end{center}
\end{table*}
This paper has provided the information-theoretic security proof of the differential-phase-shift (DPS) QKD protocol in the finite-size regime. The main analytical result is Theorem~\ref{th:mainPH}, which shows the upper bound on the number of phase errors in the finite-size regime. 
For better performance, our analysis employs Kato's inequality~\cite{kato} to upper-bound the sum of 
conditional expectations regarding the three-photon emission events. 
If we use Azuma's inequality~\cite{Azuma} instead of Kato's one, the key rate is significantly degraded. 
This is because the deviation term of Azuma's inequality scales with the square root of the number of trials, and hence if the sum of 
conditional expectations is much smaller than the number of trials, which is the case for the three-photon emission events, this inequality only gives a loose bound. 
Fortunately, however, we have revealed that Kato's inequality gives a much tighter deviation term than Azuma's one for these three-photon emission events 
and the key rate is drastically improved. As a result of our security analysis, our numerical simulation in Fig.~\ref{fig:key} 
has shown that Alice and Bob can generate a 3 Mbit secret key over 77 km for 8.3 hours under typical experimental parameters. 
Therefore, our results strongly suggest the feasibility of the DPS QKD under a realistic experimental setup.

We end with some open questions. 
In practical situations, it could be difficult 
for the assumption \ref{assV} to be satisfied, which requires that the vacuum emission probabilities are the same between both bit values. 
This issue was already solved in~\cite{mizutani_single} only in the asymptotic regime. 
Hence, it has of practical importance to reveal how the difference in these vacuum probabilities affects the key rate in the finite-size regime. 

{\it Note added.} 
After we posted the paper on the arXiv, we became aware of the independent related work~\cite{DPSeacc} that provides 
a finite-key security analysis of the DPS protocol using the entropy accumulation technique~\cite{metger}. 
This proof is valid against the most general attacks, but it requires the relativistic constraint to satisfy the 
sequential assumption, where Alice must wait to emit the $(i+1)^{\U{th}}$ pulse until she can be sure that the $(i+1)^{\U{th}}$ pulse 
will not affect Bob's $i^{\U{th}}$ measurement outcome. 
For instance, 
this can be realized by Alice sending the $(i+1)^{\U{th}}$ pulse after Bob completes the measurement of the $i^{\U{th}}$ pulse. 
The key rate per pulse $R$ of Ref.~\cite{DPSeacc} is in the order of $O(\eta)$ while our key rate is $R=O(\eta^2)$ with $\eta$ denoting the channel transmission. 
Importantly, however, our proof is free from such a relativistic constraint, which 
implies that our protocol can increase the repetition rate of the protocol as much as possible. 
Hence, even if our key rate per pulse is inferior to that in Ref.~\cite{DPSeacc}, 
our key rate {\it per second} could exceed the one of Ref.~\cite{DPSeacc} in some distance regime. 
Regarding the device models assumed in the security proofs, our proof holds even under the 
existence of source imperfections but assumes the PNR detectors, while the proof in~\cite{DPSeacc} assumes ideal 
coherent states but holds with the threshold detectors. 
We summarize in Table~\ref{table:I} the differences between our proof and the one in~\cite{DPSeacc}.

\section*{Acknowledgements}
We thank Hiroki Takesue, Toshimori Honjo, Koji Azuma, Takuya Ikuta, Hsin-Pin Lo and Guillermo Curr\'as-Lorenzo for helpful discussions. 
A.M. is supported by JST, ACT-X Grant No. JPMJAX210O, Japan. 
Y.T. is supported by the MEXT Quantum Leap Flagship Program (MEXT Q-LEAP) Grant Number JPMXS0118067394 and JPMXS0120319794, JST [Moonshot R\&D -- MILLENNIA Program] Grant Number JPMJMS2061, and the Grant-in-Aid for Scientific Research (A) No.JP22H00522 of JSPS. 
K.T. acknowledges support from JSPS KAKENHI Grant Number JP18H05237.

\begin{widetext}
\appendix
\section{Proof of Eq.~(\ref{eq:comb})}
\label{app:secrecyconddecomp}
In this appendix, we prove Eq.~(\ref{eq:comb}). 
For this, we introduce the intermediate state 
\begin{align}
\hat{\sigma}_{ABE|N_{\fin}}:=\sum_{k_A,k_B}\Pr[k_A,k_B|N_{\fin}]|k_A,k_A\rangle\langle k_A,k_A|_{AB}\otimes
\hat{\rho}^{\fin}_{E|N_{\fin}}(k_A,k_B), 
\label{eq:ideal_int}
\end{align}
and the triangle inequality of the 1-norm gives
\begin{align}
||\hat{\rho}_{ABE|N_{\fin}}^{\rm ideal}-\hat{\rho}_{ABE|N_{\fin}}^{\rm fin}||_1\le
||\hat{\rho}_{ABE|N_{\fin}}^{\rm ideal}-\hat{\sigma}_{ABE|N_{\fin}}||_1+
||\hat{\sigma}_{ABE|N_{\fin}}-\hat{\rho}_{ABE|N_{\fin}}^{\rm fin}||_1.
\label{eq:tdr}
\end{align}
We first calculate the first term as follows:
\begin{align}
&||\hat{\rho}_{ABE|N_{\fin}}^{\rm ideal}-\hat{\sigma}_{ABE|N_{\fin}}||_1\no\\
=&\leftbar\frac{1}{2^{N_{\fin}}}\sum_k|{k,k}\rangle\langle{k,k}|_{AB}\otimes\tr_{AB}(\hat{\rho}^{\fin}_{ABE|N_{\fin}})-
\sum_{k_A,k_B}\Pr[k_A,k_B|N_{\fin}]|k_A,k_A\rangle\langle k_A,k_A|_{AB}\otimes\hat{\rho}^{\fin}_{E|N_{\fin}}(k_A,k_B)\rightbar_1\\
=&\leftbar\hat{U}\left[\frac{1}{2^{N_{\fin}}}\sum_k|{k,k}\rangle\langle{k,k}|_{AB}\otimes\tr_{AB}(\hat{\rho}^{\fin}_{ABE|N_{\fin}})-
\sum_{k_A,k_B}\Pr[k_A,k_B|N_{\fin}]|k_A,k_A\rangle\langle k_A,k_A|_{AB}\otimes\hat{\rho}^{\fin}_{E|N_{\fin}}(k_A,k_B)\right]\hat{U}^{\dagger}\rightbar_1\\
=&\leftbar\frac{1}{2^{N_{\fin}}}\sum_{k}|k\rangle\langle{k}|_A\otimes\tr_{AB}(\hat{\rho}^{\fin}_{ABE|N_{\fin}})-
\sum_{k_A,k_B}\Pr[k_A,k_B|N_{\fin}]|k_A\rangle\langle k_A|_A\otimes\hat{\rho}^{\fin}_{E|N_{\fin}}(k_A,k_B)
\rightbar_1\\
=&\leftbar\hat{\rho}_{AE|N_{\fin}}^{\rm ideal}-\hat{\rho}_{AE|N_{\fin}}^{\fin}
\rightbar_1.
\label{eq:eps}
\end{align}
We obtain the first equality by substituting the definitions in Eqs.~(\ref{eq:ideal}) and (\ref{eq:ideal_int}). 
The second equality follows from the unitary-invariance property of the 1-norm. 
The third equality follows by setting the unitary operator as 
$\hat{U}=\sum_{k}\ket{k}\bra{k}_A\otimes\bigotimes_{i=1}^{N_{\fin}}\hat{X}_{B_i}^{k_i}$ with 
$\hat{X}_B$ denoting the Pauli-$X$ operator acting on system $B$. 
The final equality follows by the definitions in Eqs.~(\ref{d1}) and (\ref{d2}). 
Combining Eqs.~(\ref{eq:secrecycond}) and (\ref{eq:eps}) results in
\begin{align}
\frac{1}{2}\sum_{N_{\fin}\ge0}\Pr[N_{\fin}]||\hat{\rho}_{ABE|N_{\fin}}^{\rm ideal}-\hat{\sigma}_{ABE|N_{\fin}}||_1
\le \epsilon_s.
\label{eq:eps2}
\end{align}

Next, we calculate the second term of Eq.~(\ref{eq:tdr}) as follows:
\begin{align}
&\leftbar\hat{\sigma}_{ABE|N_{\fin}}-\hat{\rho}_{ABE|N_{\fin}}^{\rm fin}\rightbar_1\no\\
=&\leftbar\sum_{k_A,k_B}\Pr[k_A,k_B|N_{\fin}]\left(
\hat{P}[\ket{k_A,k_A}_{AB}]-\hat{P}[\ket{k_A,k_B}_{AB}]\right)
\otimes\hat{\rho}^{\fin}_{E|N_{\fin}}(k_A,k_B)
\rightbar_1\\
=&\leftbar\sum_{\substack{k_A,k_B\\k_A\neq k_B}}\Pr[k_A,k_B|N_{\fin}]
\left(
\hat{P}[\ket{k_A,k_A}_{AB}]-\hat{P}[\ket{k_A,k_B}_{AB}]
\right)
\otimes\hat{\rho}^{\fin}_{E|N_{\fin}}(k_A,k_B)
\rightbar_1\\
=&\leftbar\sum_{\substack{k_A,k_B\\k_A\neq k_B}}\Pr[k_A,k_B|N_{\fin}]
\hat{P}[\ket{k_A,k_A}]
\otimes\hat{\rho}^{\fin}_{E|N_{\fin}}(k_A,k_B)\rightbar_1+
\leftbar\sum_{\substack{k_A,k_B\\k_A\neq k_B}}\Pr[k_A,k_B|N_{\fin}]\hat{P}[\ket{k_A,k_B}]\otimes\hat{\rho}^{\fin}_{E|N_{\fin}}(k_A,k_B)
\rightbar_1\\
=&\sum_{k_A,k_B:k_A\neq k_B}\Pr[k_A,k_B|N_{\fin}]+\sum_{k_A,k_B:k_A\neq k_B}\Pr[k_A,k_B|N_{\fin}]
=2\Pr[k_A\neq k_B|N_{\fin}].
\label{eq:epc}
\end{align}
We obtain the first equality by substituting the definitions in Eqs.~(\ref{eq:ideal_int}) and (\ref{eq:act}). 
The third equality follows from $\expect{k_A,k_B|k_A,k_A}=0$ with $k_A\neq k_B$. 
The fourth equality follows by $||\hat{A}||_1=\tr(\hat{A})$ for $\hat{A}\ge0$. 
From Eqs.~(\ref{eq:correctcond}) and (\ref{eq:epc}), we have
\begin{align}
\frac{1}{2}\sum_{N_{\fin}\ge0}\Pr[N_{\fin}]\leftbar\hat{\sigma}_{ABE|N_{\fin}}-\hat{\rho}_{ABE|N_{\fin}}^{\rm fin}\rightbar_1
\le&\epsilon_c.
\label{eq:epc2}
\end{align}
Combining Eqs.~(\ref{eq:tdr}), (\ref{eq:eps2}) and (\ref{eq:epc2}) results in Eq.~(\ref{eq:comb}), which ends the proof of Eq.~(\ref{eq:comb}). 
\sq

\section{Proof of Theorem~\ref{th:correct}}
\label{app:V}
In this appendix, we prove Theorem~\ref{th:correct}. 
This theorem can be obtained by calculating the LHS of Eq.~(\ref{eq:correctcond}) as
\begin{align}
\Pr[k_A\neq k_B\wedge N_{\fin}\ge0]
&=\Pr[k_A\neq k_B\wedge N_{\fin}\ge1]\\
&\le\Pr[k_A\neq k_B\wedge H_{\ec}(\kappa_A)=H_{\ec}(\kappa^{\rec}_B)]\\
&\le\Pr[\kappa_A\neq \kappa^{\rec}_B\wedge H_{\ec}(\kappa_A)=H_{\ec}(\kappa^{\rec}_B)]\\
&=\Pr[\kappa_A\neq\kappa^{\rec}_B]\cdot
\Pr[H_{\ec}(\kappa_A)=H_{\ec}(\kappa^{\rec}_B)|\kappa_A\neq\kappa^{\rec}_B]\\
&\le 
\Pr[H_{\ec}(\kappa_A)=H_{\ec}(\kappa^{\rec}_B)|\kappa_A\neq\kappa^{\rec}_B]\\
&\le 2^{-\zeta'}. 
\end{align}
The first equality follows from $\Pr[k_A\neq k_B|N_{\fin}=0]=0$. 
The first inequality follows because $H_{\ec}(\kappa_A)=H_{\ec}(\kappa^{\rec}_B)$ is a necessary condition of $N_{\fin}\ge1$.  
Recall that $H_{\ec}$ is the universal$_2$ hash function used in verification of error correction at step~\ref{BEC}. 
The second inequality follows because $k_A\neq k_B$ leads to $\kappa_A\neq \kappa^{\rec}_B$. 
The last inequality is due to the definition of the universal$_2$ hash function, namely, 
$\Pr[H_{\ec}(x)=H_{\ec}(x')]\le\frac{1}{|\mathcal{Y}|}$ holds 
for any pair of distinct elements $x,x'\in\mathcal{X}$ when the 
universal$_2$ hash function $H_{\ec}: \mathcal{X}\to\mathcal{Y}$ is chosen uniformly at random.
\sq

\section{Proof of Theorem~\ref{th:mainPH}}
\label{app:rigo}
In this section, we prove our main result, Theorem~\ref{th:mainPH}, 
by executing statistical analysis using Azuma's and Kato's inequalities and the Chernoff bound. 
To derive this theorem from Eq.~(\ref{eq:procond}), we first employ Azuma's inequality (see Appendix~\ref{sec:azuma} for details). 
In so doing, we define the following random variable: 
\begin{align}
X^{(i)}:=&\sum^i_{p=1}\Bigg[\frac{1}{t}\left(\chi^{(p)}_{\U{ph}}-E[\chi^{(p)}_{\U{ph}}|F^{(p-1)}]\right)\no\\
&-\frac{\lambda}{1-t}\left(\chi^{(p)}_{\U{bit}}-E[\chi^{(p)}_{\U{bit}}|F^{(p-1)}]\right)
-
\frac{1}{t}\sum_{a=2,3}\left(\chi^{(p)}_{a}-E[\chi^{(p)}_{a}|F^{(p-1)}]\right)\Bigg].
\label{eq:defX}
\end{align}
Since $X^{(i-1)}$ becomes constant given $F^{(i-1)}$, 
the sequence of random variables 
$\{X^{(i)}\}_{i=0}^{N_{\det}}$ with $X^{(0)}=0$ satisfies the martingale condition defined in Eq.~(\ref{eq:martingale}), namely, 
$E[X^{(i)}|F^{(i-1)}]-X^{(i-1)}=E[X^{(i)}-X^{(i-1)}|F^{(i-1)}]=0$.  
Next, we derive the bounded difference parameter $D$ in Eq.~(\ref{eq:bdc}). 
Substituting Eq.~(\ref{eq:defX}) to $|X^{(i)}-X^{(i-1)}|$ and using 
Eqs.~(\ref{eq:ig1})-(\ref{eq:ig3}) lead to
\begin{align}
|X^{(i)}-X^{(i-1)}|
=&\Bigg|\frac{1}{t}\chi^{(i)}_{\U{ph}}-\tr[\hat{e}_{\ph}\hat{\sigma}^{F^{(i-1)}}_{AB}]
-\frac{\lambda}{1-t}\chi^{(i)}_{\U{bit}}+\lambda\cdot\tr[\hat{e}_{\bit}\hat{\sigma}^{F^{(i-1)}}_{AB}]\no\\
-&\sum_{a=2,3}\left(\frac{1}{t}\chi^{(i)}_{a}-\tr[\hat{P}_{a}\hat{\sigma}^{F^{(i-1)}}_{AB}]\right)
\Bigg|.
\end{align}
If the $i^{\U{th}}$ detected round is the sample one, Alice and Bob measure their systems to learn whether a bit error occurs or not 
(Figure~\ref{diagram} depicts the measurement in the sample round). 
In this case, a possible measurement outcome is in $\{\bit,\overline{\bit}\}$, and hence we have
$|X^{(i)}-X^{(i-1)}|\le\frac{\lambda}{1-t}+1$. 
Here, we use $|a-b|\le\max\{a,b\}$ for any $a,b\ge0$. 
On the other hand, 
if the $i^{\U{th}}$ detected round is the code one, Alice learns the weight $a$, and 
Alice and Bob measure their systems to learn whether a phase error occurs or not 
(Figure~\ref{diagram} depicts the measurement in the code round). 
In this case, a possible measurement outcome is in $\bigcup_{a=0}^3\{\ph\wedge a,\overline{\ph}\wedge a\}$, and hence we have 
$|X^{(i)}-X^{(i-1)}|\le\frac{1}{t}+\lambda+1$, where we again use $|a-b|\le\max\{a,b\}$ for any $a,b\ge0$. 
By defining 
\begin{align}
D:=\max\left\{\frac{\lambda}{1-t}+1,\frac{1}{t}+\lambda+1\right\}, 
\end{align}
Azuma's inequality in Eq.~(\ref{eq:azuma}) leads to
\begin{align}
\Pr\left[X^{(N_{\det})}\ge\sqrt{2N_{\det}D^2\ln\frac{1}{\epsilon_1}}\right]\le\epsilon_1.
\label{kyk}
\end{align}
This results in
\begin{align}
&\frac{1}{t}N_{\ph}-\frac{\lambda}{1-t}N_{\bit}-\frac{1}{t}N_{a\ge2}\no\\
\le&
\sum^{N_{\det}}_{i=1}\left(
\frac{1}{t}E[\chi^{(i)}_{\U{ph}}|F^{(i-1)}]
-\frac{\lambda}{1-t}E[\chi^{(i)}_{\U{bit}}|F^{(i-1)}]
-\frac{1}{t}\sum_{a=2,3}E[\chi^{(i)}_{a}|F^{(i-1)}]\right)+\Delta(D^2,\epsilon_1),
\label{eq:Azumainter}
\end{align}
which holds except for probability $\epsilon_1$. 
Here, we define
\begin{align}
N_{\ph}:=\sum_{i=1}^{N_{\det}}\chi^{(i)}_{\U{ph}},~
N_{\bit}:=\sum_{i=1}^{N_{\det}}\chi^{(i)}_{\U{bit}},~
N_a:=\sum_{i=1}^{N_{\det}}\chi^{(i)}_{a},~
N_{a\ge m}:=\sum_{a'=m}^3N_{a=a'}. 
\end{align}
Applying Eq.~(\ref{eq:procond}) to Eq.~(\ref{eq:Azumainter}) gives
\begin{align}
&\Pr\left[
\frac{1}{t}N_{\ph}-\frac{\lambda}{1-t}N_{\bit}-\frac{1}{t}N_{a\ge2}\le
\frac{\lambda}{t}\sqrt{
\left(\sum^{N_{\det}}_{i=1}E[\chi^{(i)}_1|F^{(i-1)}]\right)
\left(\sum^{N_{\det}}_{i=1}E[\chi^{(i)}_3|F^{(i-1)}]\right)
}+\Delta(D^2,\epsilon_1)
\right]\no\\
&\ge 1-\epsilon_1.
\label{eq:azumaresult}
\end{align}
In a similar way as we applied Azuma's inequality to the random variable in Eq.~(\ref{eq:defX}), we apply this inequality 
to the first sum of the conditional expectations in the square root. 
Importantly, as explained in Sec.~\ref{sec:upphase}, we use Kato's inequality (see Appendix~\ref{sec:kato} for details) 
to bound the second sum of the conditional expectations. By applying Azuma's and Kato's inequalities to Eq.~(\ref{eq:azumaresult}), we have 
\begin{align}
\Pr\left[
\frac{1}{t}N_{\ph}-\frac{\lambda}{1-t}N_{\bit}-\frac{1}{t}N_{a\ge2}\le
\frac{\lambda}{t}\sqrt{\left[N_{1}+\Delta(1,\epsilon_1)\right]
\cdot
\left[N_{3}+\Omega(N_{\det},N_{3},N_{3}^\ast,\epsilon_1)\right]}+\Delta(D^2,\epsilon_1)
\right]\ge1-3\epsilon_1
\end{align}
with
\begin{align}
\Omega(N_{\det},N_{3},N_{3}^\ast,\epsilon_1):=\left[b^*(N_{\det},N_{3}^\ast,\epsilon_1)+
a^*(N_{\det},N_{3}^\ast,\epsilon_1)\left(\frac{2N_{3}}{N_{\det}}-1
\right)
\right]\sqrt{N_{\det}}.
\label{yyks}
\end{align}
Here, $a^*(N_{\det},N_{3}^\ast,\epsilon_1)$ and $b^*(N_{\det},N_{3}^\ast,\epsilon_1)$ are defined in Eqs.~(\ref{ang}) and (\ref{bng}), respectively, and recall that 
$N_3^\ast$ denotes the prediction of $N_3$. 
Since $[N_{3}+\Omega(N_{\det},N_{3},N_{3}^\ast,\epsilon_1)]$ is non-decreasing against $N_3$, 
we can exploit a trivial inclusion relation $N_a\le M_a$ 
that the number $M_a$ of emitted blocks with the weight $a$ is no smaller than that of $N_a$ and obtain
\begin{align}
\Pr\left[
N_{\ph}
\le\frac{\lambda te_{\bit}N_{\sample}}{1-t}+M_{a\ge2}+
\lambda
\sqrt{[M_{1}+\Delta(1,\epsilon_1)]
[M_{3}+\Omega(N_{\det},M_{3},N_{3}^\ast,\epsilon_1)]}+t\Delta(D^2,\epsilon_1)\right]\ge 1-3\epsilon_1.
\label{eq:interA}
\end{align}
Applying the Chernoff bound, for any $\epsilon_2$ ($0\le\epsilon_2\le1$), 
\begin{align}
M_{a\ge a'}\le tq_{a'}N_{\emm}+\Gamma_{a'}
\label{UM}
\end{align}
 holds except for probability $\epsilon_2$ with 
$\Gamma_a:=\frac{1}{2}\left[-\ln\epsilon_2+\sqrt{(\ln\epsilon_2)^2-8tq_aN_{\U{em}}\ln\epsilon_2}\right]$. 
Recall that probability $q_a$ is defined in Eq.~(\ref{eq:qa}). 
Substituting the upper bound in Eq.~(\ref{UM}) to Eq.~(\ref{eq:interA}), Eq.~(\ref{eq:interA}) results in Eq.~(\ref{nge}),
which ends the proof of Theorem~\ref{th:mainPH}.

\section{Concentration inequalities}
\label{app:azumach}
In this section, we describe two concentration inequalities, Azuma's and Kato's inequalities, which are used in deriving the upper-bound on the 
number of phase errors in Sec.~\ref{sec:upphase}. 

\subsection{Azuma's inequality}
\label{sec:azuma}
\begin{theorem}
\label{th:azuma}
(Azuma's inequality~\cite{Azuma})
For $n\in\mathbb{N}$, 
let $\{X^{(i)}\}_{i=0}^n$ be a sequence of random variables with $X^{(0)}=0$, and 
$\{F^{(i)}\}_{i=0}^n$ be a filtration with $F^{(i)}$ identifying the random variables including $\{X^{(0)},...,X^{(i)}\}$. 
The sequence of random variables $\{X^{(i)}\}_{i=0}^n$ satisfies the martingale condition
\begin{align}
E[X^{(i)}|F^{(i-1)}]=X^{(i-1)}
\label{eq:martingale}
\end{align}
for any $i\in\{1,...,n\}$. 
Also, the difference sequence $Y^{(i)}:=X^{(i)}-X^{(i-1)}$ satisfies the bounded difference condition, namely, there exists a positive constant $D>0$ such that 
for any $i\in\{1,...,n\}$
\begin{align}
|Y^{(i)}|\le D
\label{eq:bdc}
\end{align}
holds. Then, for any $n\in\mathbb{N}$ and any $\epsilon$ with $0\le\epsilon\le1$, 
\begin{align}
\Pr\left[X^{(n)}\ge \sqrt{2nD^2\ln\frac{1}{\epsilon}}\right]\le\epsilon.
\label{eq:azuma}
\end{align}
\end{theorem}

\subsection{Kato's inequality}
\label{sec:kato}
We explain how to apply Kato's inequality~\cite{kato} to derive an upper bound on 
$\sum^{N_{\det}}_{i=1}E[\chi^{(i)}_3|F^{(i-1)}]$ in Eq.~(\ref{eq:azumaresult}). 
Kato's inequality states that for any $a\in\mathbb{R}$ and any $b\ge|a|$,
\begin{align}
\Pr\left[
\sum^{N_{\det}}_{i=1}E[\chi^{(i)}_3|F^{(i-1)}]
\ge N_3+\left\{b+a\left(\frac{2N_3}{N_{\det}}-1\right)\right\}\sqrt{N_{\det}}
\right]
\le\exp\left[-\frac{2b^2-2a^2}{\left(1+\frac{4a}{3\sqrt{N_{\det}}}\right)^2}\right]
\label{eq:ka}
\end{align}
with $N_3=\sum_{i=1}^{N_{\det}}\chi_3^{(i)}$. 
To fix $a$ and $b$, we consider minimizing the deviation term $\left[b+a\left(\frac{2N_3}{N_{\det}}-1\right)\right]\sqrt{N_{\det}}$ 
given the failure probability [RHS of Eq.~(\ref{eq:ka})] being $\epsilon$ ($0\le\epsilon\le1$). 
But, Alice and Bob do not know the true value of $N_3$ even after running the protocol. 
Therefore, we make prediction $N_3^\ast$ of $N_3$, and using this prediction we solve the following optimization problem: 
\begin{align}
&\min~~~~\left[b+a\left(\frac{2N_3^\ast}{N_{\det}}-1\right)\right]\sqrt{N_{\det}}\\
&\U{s.t.}~~~~\exp\left[-\frac{2b^2-2a^2}{\left(1+\frac{4a}{3\sqrt{N_{\det}}}\right)^2}\right]=\epsilon\\
&~~~~~~~~~b\ge|a|.
\end{align}
This problem is analytically solved in~\cite{npj21} as $a'(N_{\det},N_3^\ast,\epsilon)$ and $b'(N_{\det},N_3^\ast,\epsilon)$ with 
\begin{align}
a'(n,m,\epsilon):=
\frac{216\sqrt{n}m(n-m)\ln\epsilon-48n^{\frac{3}{2}}(\ln\epsilon)^2
+27\sqrt{2}(n-2m)\sqrt{-n^2(\ln\epsilon)[9m(n-m)-2n\ln\epsilon]}}
{4(9n-8\ln\epsilon)[9m(n-m)-2n\ln\epsilon]
}
\end{align}
\begin{align}
b'(n,m,\epsilon):=
\frac{\sqrt{18a'(n,m,\epsilon)^2n-[16a'(n,m,\epsilon)^2+24a'(n,m,\epsilon)\sqrt{n}+9n]\ln\epsilon}}{3\sqrt{2n}}
\end{align}
under $N_3^\ast<N_{\det}/2$. 
These $a'$ and $b'$ are the optimal values of $a$ and $b$ when $N_3^\ast=N_3$ and could be near optimal when 
$N_3^\ast$ is close to $N_3$. 
To make $N_3+\left[b+a\left(\frac{2N_3}{N_{\det}}-1\right)\right]\sqrt{N_{\det}}$ non-decreasing against $N_3$, we set 
$a=a^\ast(N_{\det},N_3^\ast,\epsilon)$ and $b=b^\ast(N_{\det},N_3^\ast,\epsilon)$ with
\begin{align}
a^\ast(n,m,\epsilon):=\max\left\{-\frac{\sqrt{n}}{2},a'(n,m,\epsilon)\right\}
\end{align}
\begin{align}
b^\ast(n,m,\epsilon):=
\frac{\sqrt{18a^\ast(n,m,\epsilon)^2n-[16a^\ast(n,m,\epsilon)^2+24a^\ast(n,m,\epsilon)\sqrt{n}+9n]\ln\epsilon}}{3\sqrt{2n}}.
\end{align}
Substituting $a=a^\ast(N_{\det},N_3^\ast,\epsilon)$ and $b=b^\ast(N_{\det},N_3^\ast,\epsilon)$ to Eq.~(\ref{eq:ka}), we obtain 
\begin{align}
\sum^{N_{\det}}_{i=1}E[\chi^{(i)}_3|F^{(i-1)}]
\le N_3+\left[b^\ast(N_{\det},N_3^\ast,\epsilon)+a^\ast(N_{\det},N_3^\ast,\epsilon)\left(\frac{2N_3}{N_{\det}}-1\right)\right]\sqrt{N_{\det}}.
\end{align}
This upper bound has a free parameter $N^\ast_{3}$, which is the prediction of $N_{3}$ and can be freely chosen under 
$N^\ast_{3}<N_{\det}/2$. For this, we set 
\begin{align}
N_3^\ast=\min\left\{tq_3N_{\U{em}}+\Gamma_3,\lfloor (N_{\det}-1)/2 \rfloor\right\}, 
\end{align}
whose first element is an upper bound on $N_3$ in Eq.~(\ref{UM}). 
\end{widetext}


\begin{thebibliography}{99}
\bibitem{footnote1}
Note that $F^{(0)}=\{\emptyset,\Omega\}$ and $F^{(N_{\det})}=2^\Omega$, and 
$F^{(i)}\subseteq F^{(j)}$ with $i\le j$ means that the information of the measurement outcomes is updated.
\bibitem{footnote2}
Note that this statement is directly obtained from the assumptions on the light source in Sec.~\ref{ass:A}.
\end{thebibliography}
\end{document}